\documentclass[final,12pt]{msml2022}

\usepackage{url}            
\usepackage{booktabs}       
\usepackage{float}
\usepackage{xcolor}
\usepackage{amsmath,amsfonts,bm}
\usepackage{mathtools}
\usepackage{algorithm}
\usepackage{algorithmic}

\usepackage{hyperref}
\definecolor{mydarkblue}{rgb}{0,0.08,0.45}
\hypersetup{
  unicode=true,          pdftoolbar=true,        pdfmenubar=true,        pdffitwindow=false,     pdfstartview={FitH},    pdftitle={My title},    pdfauthor={Author},     pdfsubject={Subject},   pdfcreator={Creator},   pdfproducer={Producer}, pdfkeywords={keyword1, key2, key3}, pdfnewwindow=true,      colorlinks=true,       linkcolor=mydarkblue,          citecolor=mydarkblue,        filecolor=mydarkblue,      urlcolor=mydarkblue           }

\numberwithin{equation}{section}
\usepackage{lipsum}  
\newcommand*\samethanks[1][\value{footnote}]{\footnotemark[#1]}

\usepackage{physics}

\newcommand{\params}{{\bm\theta}}
\newcommand{\ba}{{\bm\alpha}}
\newcommand{\bt}{{\bm\tau}}
\newcommand{\psii}{{\psi_{\text{init}}}}
\newcommand{\pgs}{{\psi_{\text{GS}}}}

\usepackage{indentfirst}
\renewcommand{\eqref}[1]{(\ref{#1})}
\usepackage{enumitem,amssymb}
\newlist{todolist}{itemize}{2}
\setlist[todolist]{label=$\square$}
\usepackage{pifont}
\newcommand{\cmark}{\textcolor{tgreen}{\ding{51}}}\newcommand{\xmark}{\textcolor{tred}{\ding{55}}}\newcommand{\cxmark}{\textcolor{torange}{\ding{51}}{\small\textcolor{torange}{\kern-0.7em\ding{55}}}}

\usepackage{multirow}
\usepackage{colortbl}
\newcommand{\midsepremove}{\aboverulesep = 0mm \belowrulesep = 0mm \extrarowheight = 0.85ex}
\midsepremove
\newcommand{\midsepdefault}{\aboverulesep = 0.605mm \belowrulesep = 0.984mm \extrarowheight = 0mm}
\midsepdefault

\definecolor{tblue}{HTML}{1f77b4}
\definecolor{torange}{HTML}{ff7f0e}
\definecolor{tgreen}{HTML}{2ca02c}
\definecolor{tred}{HTML}{d62728}

\usepackage{xspace}
\usepackage{siunitx}
\renewcommand{\cite}{\citep} 
\renewcommand{\citet}{\citep} 
\usepackage{soul}
\newcommand{\IfThen}[2]{\STATE \algorithmicif\ #1\ \algorithmicthen\ #2}

\newcommand{\agentname}{MCTS-QAOA\xspace}

\title[\agentname]{Monte Carlo Tree Search based Hybrid Optimization \\ of Variational Quantum Circuits}
\usepackage{times}

\msmlauthor{\Name{Jiahao Yao}\thanks{ J.Y. \& H.L. contributed equally} \Email{jiahaoyao@berkeley.edu}\\
 \addr  Department of Mathematics, University of California, Berkeley\\ Berkeley,  CA 94720, USA
  \AND
  \Name{Haoya Li}\samethanks[1] \Email{lihaoya@stanford.edu} \\
 \addr Department of Mathematics, Stanford University, Stanford, USA
 \AND
  \Name{Marin Bukov} \\
 \addr Department of Physics, St.~Kliment Ohridski University of Sofia\\ 5 James Bourchier Blvd, 1164 Sofia, Bulgaria \\
 Max Planck Institute for the Physics of Complex Systems, N\"othnitzer Str.~38, 01187 Dresden, Germany
 \AND
  \Name{Lin Lin}  \\
 \addr Department of Mathematics, University of California, Berkeley\\
Computational Research Division, Lawrence Berkeley National Laboratory\\
Challenge Institute for Quantum Computation, University of California, Berkeley\\
Berkeley, CA 94720, USA
\AND
 \Name{Lexing Ying} \\
 \addr Department of Mathematics, Stanford University, Stanford, USA
}

\makeatletter
 \let\Ginclude@graphics\@org@Ginclude@graphics
\makeatother

\begin{document}
\maketitle

\begin{abstract}

  Variational quantum algorithms stand at the forefront of simulations on near-term and future fault-tolerant quantum devices. While most variational quantum algorithms involve only continuous optimization variables, the representational power of the variational ansatz can sometimes be significantly enhanced by adding certain discrete optimization variables, as is exemplified by the generalized quantum approximate optimization algorithm (QAOA). However, the hybrid discrete-continuous optimization problem in the generalized QAOA poses a challenge to the optimization.  We propose a new algorithm called \agentname, which combines a Monte Carlo tree search method with an improved natural policy gradient solver to optimize the discrete and continuous variables in the quantum circuit, respectively. We find that \agentname has excellent noise-resilience properties and outperforms prior algorithms in challenging instances of the generalized QAOA.

\end{abstract}

\section{Introduction}

Quantum computing provides a fundamentally different way for solving a variety of important problems in scientific computing, such as finding the ground state energy in computational chemistry, and the MaxCut problem in combinatorial optimization.
Variational quantum circuits are perhaps the most important quantum algorithms on near term quantum devices~\citep{preskill2018quantum}, mainly due to the tunability and the relatively short circuit depth~\citep{CerezoArrasmithBabbushEtAl2021}, as exemplified by the variational quantum eigensolver (VQE) \citep{peruzzo2014variational,McCleanRomeroBabbushEtAl2016} and the quantum approximate optimization algorithm (QAOA)~\citep{farhi2014quantum}. A common thread in these algorithms is to  variationally optimize a parameterized quantum circuit using classical methods to obtain an approximate ground state. For instance, in combinatorial optimization, QAOA encodes the classical objective function  into a quantum Hamiltonian, and constructs a quantum circuit with a set of two alternating quantum gates. The continuous adjustable parameters are the duration or phases of the gates.

For quantum many-body problems, the expressivity of the QAOA ansatz may be limited: the exponentially large (in the number of qubits) Hilbert space may not be efficiently navigated by the dynamics generated by the alternating gate sequence. This can lead to circuit depths that grow with the system size~\citep{ho2019efficient}, or render the target ground state outside the scope of accessible states altogether, thus fundamentally precluding its preparation. To address these problems, various versions of a generalized QAOA ansatz have been presented in recent works~\citet{zhu2020adaptive, yao2020reinforcement, chandarana2021digitized}, where additional control Hamiltonians are used to generate the variational circuits.
In general, these Hamiltonians are tailored to the many-body system whose ground state we seek to prepare, and the extended Hamiltonian pool is often constructed using ideas from variational counter-diabatic (CD) driving~\citep{sels2017minimizing}.
When the optimization of the parameterized circuit is performed successfully, the generalized ansatz produces a closer approximation to the ground state than the original alternating QAOA ansatz. The generalized QAOA may also significantly reduce the total protocol duration $T$ and therefore the depth of the quantum circuit while giving a high fidelity with respect to the ground state~\cite{yao2020reinforcement}.

\begin{figure}[t!]
  \centering
  \includegraphics[width=1.0\textwidth]{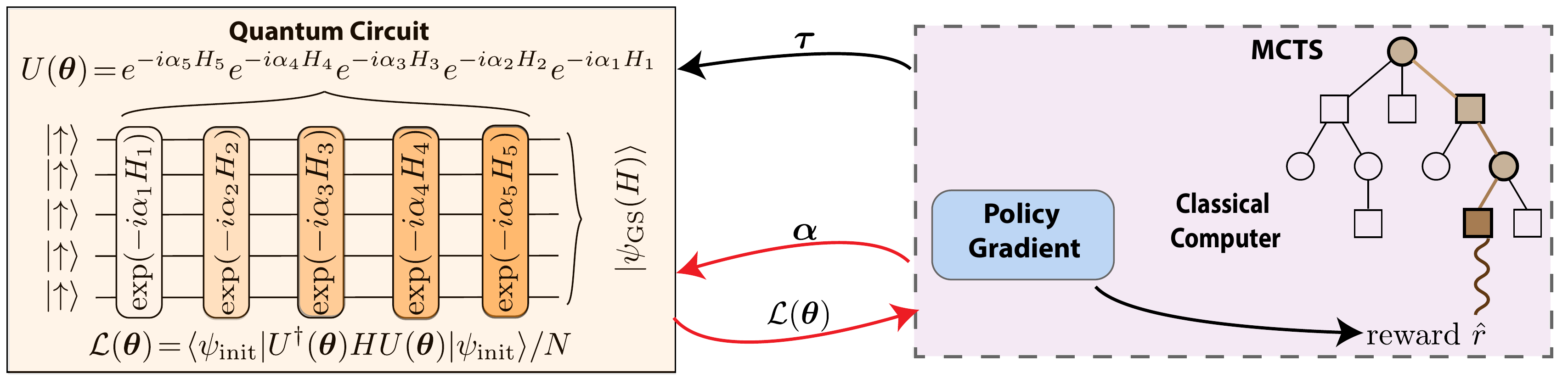}
  \caption{
    \small \textbf{The schematics of \agentname}: MCTS provides promising paths for the discrete optimization search; the inner loop (highlighted in red) Policy Gradient (PG) solver evaluates the discrete sequence in a noise-robust way; the reward obtained is then propagated back through the search tree and used to improve the tree policy.
  }
  \label{fig:concept}
\end{figure}

However, the ansatz  of the generalized QAOA also results in a more challenging optimization problem. The original QAOA only involves optimization of continuous parameters. The generalized QAOA ansatz, in contrast, leads to a hybrid optimization problem that involves both the discrete variables (the choice of quantum gates) and the continuous variables (the duration of each gate). To solve this hybrid optimization problem, we propose a novel algorithm that combines the Monte Carlo Tree Search (MCTS) algorithm~\citep{coulom2006efficient, browne2012survey, abramson2014expected, silver2016mastering, silver2017mastering} -- a powerful method in exploring the discrete sequence, with an improved noise-robust natural policy gradient solver for the continuous variables of a fixed gate sequence.

\noindent\textbf{Contributions:}
\begin{itemize}
  \item We propose the \agentname algorithm which combines the MCTS algorithm and a noise-robust policy gradient solver. We show that it is not only efficient in exploring the quantum gate sequences but also robust in the presence of different types of noise.

  \item The proposed \agentname{} algorithm produces accurate results for problems that appear difficult or infeasible for previous algorithms based on the generalized QAOA ansatz, such as RL-QAOA~\citep{yao2020noise}. In particular, \agentname{} shows superior performance in the large protocol duration regime, where the hybrid optimization becomes challenging.

  \item In order for the \agentname{} algorithm to produce reliable optimal results, it is crucial that the inner loop solver finds the optimal continuous variables with high accuracy. Compared to the original PG-QAOA solver introduced in~\citet{yao2020policy}, we improve the inner loop solver with entropy regularization and the natural gradient method, and implement it in Jax~\citep{jax2018github}, which offers more accurate, stable, and efficiently computed solutions during the continuous optimization.

  \item For the physics models considered in this paper, we observe that there can be many ``good'' gate sequences. This means that for a large portion of gate sequences, the energy ratio obtained is not far away from the optimal energy ratio obtainable with the generalized QAOA ansatz, given that the continuous variables are solved with high quality. This phenomenon has not been recorded in the literature to the best of the authors' knowledge.
\end{itemize}

\noindent\textbf{Related works:}

\noindent\textbf{Quantum control and variational quantum eigensolver}: Traditional optimal quantum control methods, often used in prior works, are GRAPE~\citep{khaneja2005optimal} and CRAB~\citep{caneva2011chopped}. More recently, success has been seen by the combination of traditional methods with machine learning ~\citep{schafer2020differentiable,wang2020bayesian,sauvage2019optimal,fosel2020efficient,nautrup2019optimizing,albarran2018measurement,sim2020adaptive,wu2020end,wu2020active, abhinav2020natural, dalgaard2022predicting}, and especially reinforcement learning~\citep{niu2019universal, fosel2018reinforcement, august2018taking,  porotti2019coherent,wauters2020reinforcement,yao2020policy,sung2020towards, chen2013fidelity, bukov2018reinforcement, bukov2018day, sordal2019deep, bolens2020reinforcement, dalgaard2020global, metz2022self}). Among them, Variational quantum eigensolver or VQE~\citep{cerezo2021variational, tilly2021the} provides a general framework applicable on noisy intermediate-scale quantum (NISQ) devices \citep{preskill2018quantum} to variationally tune the circuit parameters and improve the approximation. In the fault tolerant setting, there are also possibilities of error mitigation via the variational quantum optimization~\citep{sung2020exploration, arute2020hartree}.

QAOA~\citep{farhi2014quantum} can be viewed as a specific variational quantum algorithm, and can be extended to the generalized QAOA ansatz~\citet{zhu2020adaptive, yao2020reinforcement, chandarana2021digitized}. Prior works optimize the generalized QAOA greedily and progressively for each circuit layer or end-to-end as a large autoregressive network. The present work differs from these methods; we take advantage of the MCTS structure and formulate the problem as a two-level optimization.

\noindent\textbf{MCTS and RL}: Monte Carlo tree search (MCTS) has been one major workhorse behind the recent breakthrough of reinforcement learning algorithm, especially AlphaGo algorithms and variants~\citep{silver2016mastering, silver2017mastering, silver2018general, schrittwieser2020mastering, ye2021mastering}. MCTS~\citep{browne2012survey, guo2014deep} makes use of a discrete hierarchical structure to figure out a better exploration in high dimensional search problems. While it is typically applied to discrete search, it has also been used in the continuous setting~\citep{wang2020learning}, where the partition space of the whole space is viewed as branching of the tree. In the context of quantum computing, applications of MCTS have been recently emerged such as the Quantum Circuit Transformation~\citep{zhou2020a}, the quantum annealing schedules~\citep{chen2020optimizing}, and the quantum dynamics optimization~\citep{dalgaard2020global}.

Further related works in  hybrid optimization,  counter-diabatic driving methods, and architecture search can be found in Appendix~\ref{app:related}.

\section{Generalized QAOA ansatz}

The generalized QAOA ansatz \citep{yao2020reinforcement} constructs a variational quantum circuit via the composition of a sequence of parameterized unitary operators:
\begin{equation}
  \label{eq:U_ansatz}
  U(\params)\!=\!\prod_{j=1}^q U(\tau_j, \alpha_j) \!= \!\prod_{j=1}^q\exp(-i\alpha_jH_{\tau_j}).
\end{equation}
Here the circuit parameters $\params\!=\!(\ba, \bt)$ contain two components: i) the \textit{discrete} variables $\bt=(\tau_1, \tau_2, \ldots, \tau_q)$ define a sequence of Hamiltonians with length $q$, while ii) the \textit{continuous} variables $\ba\!=\!\{\alpha_j\}_{j=1}^q$ represent the duration that each corresponding gate is applied for. It is further assumed that each Hamiltonian $H_{\tau_j}$ is selected from a fixed Hamiltonian pool $\mathcal{A}\!=\!\{H_1, H_2, \cdots, H_{\abs{\mathcal{A}}}\}$, and consecutive gates are not repeated,~i.e.,~$\tau_j\not=\tau_{j+1}, ~1\leq j\leq q-1$. The total number of possible sequences is thus $|\mathcal{A}|(|\mathcal{A}|-1)^{q-1}$, which grows exponentially with $q$, rendering exhaustive search intractable.

After applying the circuit to an initial quantum state $\ket{\psi_{\text{init}}}$, one obtains the final quantum state $\ket{\psi} = U(\params)\ket{\psi_{\text{init}}}$. To prepare a high quality approximation of the ground state $\ket{\psi_\text{GS}}$ of the target Hamiltonian $H$, the continuous and discrete variables are solved for by minimizing the following objective function:
\begin{equation}
  \label{eq:loss}
  \mathcal L(\params)\!=\!E(\params)/N \!=\! \langle\psi_{\text{init}}|U^\dagger(\params) H U(\params) |\psi_{\text{init}}\rangle/N.
\end{equation}
Note that the energy $E$ in the objective function is divided by the number of particles $N$ in the physical model, e.g.~the number of qubits. This scaled objective function has a well-behaved limit when increasing the number of qubits, as required for larger-scale computations. Here, the energy function $E(\params)$ is always lower-bounded by the ground state energy $E_\text{GS}\!=\!\langle\psi_\text{GS}|H|\psi_\text{GS}\rangle$. It is also worth noticing that the quantum states $|\psi\rangle$ are unknown to the optimization algorithm (they cannot be measured), which increases the difficulty of the optimization algorithm.

\section{Reinforcement learning setup}\label{sec:RLsetup}

After defining the optimization problem posed by the generalized QAOA, let us briefly cast it within the RL framework.

\subsection{Quantum constraints on the RL environment}

Beyond classical physics, quantum mechanics imposes counterintuitive constraints on the state and reward spaces, which need to be embedded in a realistic RL environment.

First, the quantum state (or wavefunction) is not a physical observable by itself, and inference of the information of the full quantum state from experiments (called quantum state tomography) can require exponential resources. This fact is intimately related to the expected superior performance of quantum computers against their classical counterparts on certain tasks. To embed this quantum behavior into our environment simulator, we define the RL state as the sequence of actions applied~\citep{bukov2018reinforcement} rather than the quantum state.
Starting from a fixed initial state, the quantum state is uniquely determined (though still unmeasurable) by the Hamiltonian sequence applied.

Second, (strong) quantum measurements lead to a collapse of the quantum wavefunction. This means that, once a measurement has been performed, the state itself is irreversibly lost. Therefore, a second constraint for our quantum RL environment is the sparsity of rewards. Indeed, only after the RL episode comes to an end, can we measure the energy and obtain the reward. In Sec.~\ref{sec:method}, we exploit this fact to introduce MCTS into the algorithm which does not evaluate the protocol $\bt$ during the construction of it. As a result, the evaluation is delegated to the noise-robust PG-QAOA solver.

\subsection{The reinforcement learning environment}
In the language of reinforcement learning (RL), the choice of quantum gates corresponds to the action of the learner, and the quantum circuit is completed after $q$ actions, which marks the end of the RL session/episode. The reward signal is provided by the inner loop solver which aims to compute the lowest possible energy that can be reached by the fixed chosen gate sequence. To be more specific, the action space $\mathcal{A} = \{H_j:1\leq j\leq |\mathcal{A}|\}$ is a set of Hamiltonians; the state space $\mathcal{S} = \{(\tau_1, \tau_2, \ldots, \tau_t): \tau_j\in \mathcal{A}, 0\leq j\leq t, 1\leq t \leq q\}$ is the set of sequences of Hamiltonians with length no larger than $q$. In particular, a session always starts with the empty sequence $s_0$, and ends with a state given by a Hamiltonian sequence of length $q$. When $s_t = (\tau_1, \tau_2, \ldots, \tau_t)$ is not a terminal state, i.e., $t\!<\! q$, the next state $s_{t+1}$ is obtained by appending the $(t\!+\! 1)$-th action $\tau_{t+1}$ at the end of $s_t$, i.e., $s_{t+1} = (\tau_1, \tau_2, \ldots, \tau_t, \tau_{t+1})$.

The reward $r(s)$ only depends on the state $s$, and it is set as $0$ whenever $s$ is not a terminal state. As explained in the previous section, this implements the physical constraint reflecting the inability to perform a strong quantum measurement without destroying the quantum state. When $s$ is a terminal state $\bt=(\tau_1, \tau_2, \ldots, \tau_q)$, we define
\begin{equation}
  \label{eq:reward_signal}
  r(s) = r(\bt) = -\underset{\alpha}{\min} ~E(\{\alpha_j\}_{j=1}^q, \bt)/N,
\end{equation}
where
$\{\alpha_j\}_{j=1}^q$ are the duration obtained by the inner loop continuous optimizer,
and the energy $E$ is defined in \eqref{eq:loss}.

\begin{algorithm}[!t]
  \caption{ \agentname{} }
  \label{alg:Q-MCTS}
  \begin{algorithmic}[1]
    \REQUIRE UCB bound coefficient $c$,  number of outer loop iterations $T_{\mathrm{iter}}$, number of random initialization $T_{\mathrm{init}}$. \\
    \STATE Initialize the Monte Carlo tree.
    \FOR {$t=1,..,T_{\mathrm{iter}}$}
    \STATE Pick a node according to the tree policy $\pi_\mathrm{tree}$, cf.~Eq.~\eqref{ucb1_update}, using the UCB bound with parameter $c$.
    \IF{the tree node is not the terminal state}
    \STATE Randomly roll out from the current tree node to obtain a terminal state $\bt_t$.
    \ENDIF
    \FOR {$i=1,..,T_{\mathrm{init}}$}
    \STATE Run natural policy gradient method (see Algorithm~\ref{alg:inner}) to obtain the estimated reward $r_t^{[i]}$.
    \ENDFOR
    \STATE Choose the best gate sequence durations according to the maximum reward $\hat{r}_t=\max_{i} r_t^{[i]}$ across different random intialization of policy gradient.
    \STATE Back-propagate the reward $\hat{r}_t$ from the node up to the root and update the statistics ($Q, N$) on each node.
    \ENDFOR
  \end{algorithmic}
  \vspace{0.5em}
\end{algorithm}

\section{Monte Carlo tree search with improved policy gradient solver}\label{sec:method}

In this section, we introduce~\agentname{}, an algorithm that solves the hybrid optimization problem defined by the generalized QAOA ansatz, using a combination of MCTS and an improved policy gradient solver. In the combined algorithm, MCTS serves as the solver for the outer optimization problem: it is used to search for high quality gate sequences $\bt$. At the same time, we design an improved policy gradient solver to produce the optimal gates duration $\ba$ for the discrete sequence provided by MCTS. Finally, the outcome of the evaluation is propagated back through the nodes of the MC tree to improve the tree policy before the next iteration.

\subsection{Discrete optimization: Monte Carlo tree search}

\agentname{} strikes an efficient balance between exploration and exploitation of the RL states,
by leveraging the statistics recorded in a search tree. Each node of this tree corresponds to a state $s$; the child nodes denote all possible states $s'$ following the state $s$. For the problem considered in this paper, trajectories are loop-free,
since each child state $s'$ has one more action attached than its parent state $s$. Thus, we refer to a given node by its corresponding state. In particular, the root node corresponds to the empty state $s_0$, which has $|\mathcal{A}|$ children, one for each action; any other non-terminal state $s$ has $|\mathcal{A}|\!-\! 1$ children, reflecting the constraint that no action can follow itself, and a terminal state has none. During the search process, each node keeps track of the statistics of two quantities:
i) $N(s, a)$ counts the selection of action $a$ at state $s$;
ii) $Q(s, a)$ is the expected reward after taking action $a$ at state $s$.
Intuitively, the average $Q(s, a) / N(s,a)$ is an estimate of how promising a child node is. Finally, a node $s$ is called fully expanded, if all its children are visited in the search, i.e., if $N(s, a)\!\geq\! 1$ for all $a\!\in\! \mathcal{A}$; otherwise, $s$ is called an expandable node, and is the focus of exploration.

In each MCTS iteration, the tree and the node statistics are updated as follows:
\vspace{-10pt}
\begin{enumerate}
  \item \textit{Forming a search path.} Starting from the root node, if the current node is fully expanded, then one of its children is chosen according to the following Upper Confidence Bound (UCB)~\citep{auer2002finite}:
        \begin{equation}
          \label{ucb1_update}
          \pi_\mathrm{tree}(s)\! =\! \underset{a \in \mathcal{A}}{\text{arg}\max}\! \left( \frac{Q(s, a)}{N(s, a)} +  c \sqrt{\frac{2 \log N(s)}{N(s, a)}} \right),
        \end{equation}
        until reaching a terminal state or an expandable node; here $\pi_\mathrm{tree}(s)$ denotes the tree policy.
        Then an unvisited child of the current node is chosen at random, unless the current node is a terminal state. After that, a simulation is rolled out with a uniform policy until reaching a terminal state.

  \item \textit{Evaluation and backup.} The reward $\hat{r}$\footnote{In order to distinguish the estimated reward from the true reward $r(s)$ in the presence of noise, we denote the estimated reward as $\hat{r}$.} of the terminal state is evaluated by the inner loop solver and the tree statistics are updated using $Q(s, a) \leftarrow Q(s, a) + \hat{r}, N(s, a) \leftarrow N(s, a) + 1 $ for each visited edge $(s, a)$.
\end{enumerate}

For the generalized QAOA ansatz, the real challenge lies in the evaluation step. On the one hand, the overall minimization of the energy depends on the potential of the trajectory selected by the MCTS, whose role is to find the optimal trajectory sequence. On the other hand, if the accuracy of the evaluation is low, then the searching process can be stuck at a severely suboptimal solution. Similarly, if the evaluation is not efficient enough, then the benefit obtained by using quantum computation strategy will also be lost. And last but not least, if the evaluation results are not robust to noise, then the algorithm can hardly be carried out on quantum devices. Hence, the inner loop solver used to implement the evaluation must be able to efficiently offer high accuracy results while being robust to different kinds of noise. The above considerations refer to the generic case; in practice, the optimization dynamics of the algorithm is set by the properties of the optimization landscape.

\subsection{Continuous optimization: natural policy gradient solver}
\label{sec:inner}

For each terminal state $\bt = (\tau_1, \tau_2, \ldots, \tau_q)$ reached in the MCTS process, an inner loop solver is invoked to produce the optimal duration $\ba\!=\!\{\alpha_j\}_{j=1}^q$ and the reward $ -E(\{\alpha_j\}_{j=1}^q, \bt)/N$ which are then back-propagated through the tree to update the tree statistics.
In order to ensure that the duration obtained has a practical magnitude and to allow for a fair comparison between algorithms, we further assume that the total duration of all gates is fixed as $T$, which can be seen as a protocol for the circuit depth.

The continuous optimization problem for the inner-loop solver in the reward-evaluation step is thus
\begin{equation}
  \label{eq:sim_problem}
  \min_{\{\alpha_j\}_{j=1}^q} \; \!\!\!\!\left\{ E(\{\alpha_j\}_{j=1}^q, \bt)\! :\! \sum_{j=1}^q \alpha_j \!=\! T;\ 0 \leq \alpha_j \leq T \right\}.
\end{equation}
In order to avoid using explicit derivatives of the energy $E$, we instead
optimize the expectation of the energy $E$ over a parameterized probability distribution of $\ba$; this is also crucial to to make the algorithm resilient to noise. More specifically, we set $\alpha_j = \frac{T\tilde\alpha_j}{\sum_k\tilde\alpha_k}$ to ensure the constraints on $\alpha_j$, where $\tilde{\alpha}_j$ is a random variable drawn from the sigmoid Gaussian distribution $\mathcal{SN}(\mu_j, \sigma_j)$\footnote{$\mathcal {SN}(\mu, \sigma)$ denotes the sigmoid Gaussian distribution with parameters $\mu$ and $\sigma$, i.e., the distribution of the Gaussian random variable $\mathcal{N}(\mu, \sigma)$ under the sigmoid transformation. It is also called the logit-normal distribution}. It can be parameterized as $\tilde{\alpha}_j = \mathfrak{g}(\delta_j)$, where $\delta_j\sim\mathcal{N}(\mu_j, \sigma_j)$ is a Gaussian random variable and $\mathfrak{g}(x) = \frac{1}{1+\exp(-x)}$ is the sigmoid function.
Adding a Shannon entropy regularizer to the total expected reward we obtain the regularized objective function:
\begin{equation}
  \mathcal{J}(\{\mu_{j}, \sigma_{j}\}_{j=1}^q)\!=\! \mathbb {E}_{\substack{\delta_j\sim \mathcal {N}(\mu_{j}, \sigma_{j}) }}
  \left[R(\bm\delta)\right] + \beta^{-1}_{{S}}\sum_{j=1}^q \log\sigma_j,
  \label{eqn:pg-loss}
\end{equation}
which is maximized over the parameters $\{\mu_{j}, \sigma_{j}\}_{j=1}^q$. Here $R(\bm\delta) = -E\left(\left\{\frac{T\mathfrak{g}(\delta_j)}{\sum_k\mathfrak{g}(\delta_k)}\right\}_{j=1}^q,\bt\right)/N$, and $\beta^{-1}_{{S}}$ denotes the temperature, which controls the trade-off between exploration and exploitation: higher temperature $\beta^{-1}_{S}$ leads to a larger weight on the entropy term, and thus encourages exploration, while smaller $\beta^{-1}_{S}$ reduces exploration. The entropy term $\sum_{j=1}^q \log\sigma_j$ can be derived from the definition of Shannon entropy, cf.~Appendix~\ref{app:npg}.

The inner loop solver is then constructed with a natural policy gradient (NPG) method applied to the regularized objective function $\mathcal J$ using the natural gradient direction $F^{-1}\nabla\mathcal{J}$, where $F$ is the Fisher information matrix for the joint distribution of $\{\delta_j\}_{j=1}^q$ and $\nabla \mathcal J$ is the gradient of $\mathcal J$ with respect to the parameters. This procedure is different from the solver established in PG-QAOA~\citep{yao2020policy}, where the standard gradient is used to update the parameters and no regularization is used.
Using independent standard normal variables $\xi_j$, the natural gradient direction can be approximated by unbiased estimators:
\begin{equation}\label{eq:NPGupdate}
  F_j^{-1}\begin{bmatrix}
    \frac{\partial \mathcal{J}}{\partial \mu_j} \\\frac{{\partial\mathcal{J}}}{\partial \log\sigma_j}
  \end{bmatrix}\approx
  \begin{bmatrix}
    \sigma_jR(\bm\delta)\xi_j \\
    \frac{1}{2}R(\bm\delta)(\xi_j^2-1) + \frac{1}{2}\beta^{-1}_{{S}}
  \end{bmatrix},
\end{equation}
where $\delta_j = \mu_j+\sigma_j\xi_j$ and $F_j$ is the $j$-th $2$-by-$2$ diagonal block of the Fisher information matrix, since $F$ is a block diagonal matrix, cf.~Appendix~\ref{app:npg}. In practice, we update $\log\sigma$ instead of $\sigma$ to ensure the positivity of $\sigma$, and we use the average of the unbiased estimators in \eqref{eq:NPGupdate} within a batch of size $M$ to give the approximation of the natural gradient direction.

The first term in the objective function $\mathcal{J}$ can also be viewed as a smoothed reward function obtained with Gaussian perturbation. The parameter $\{\sigma_{j}\}_{j\!=\!1}^q$ determines the distance between $\mathcal{J}(\mu_{j}, \sigma_{j})$ and $E(\mu_{j})$ \citep{nesterov2017random}. If $\sigma$ is too large, then $\mathcal{J}$ is far from $E$, and yields suboptimal solutions of $\mu_{j}$ since too much details are lost after the Gaussian smoothing. To avoid this, we propose to use a tempering technique (see for example \citet[Sec.~5]{klink2020self, abdolmaleki2018maximum, haarnoja2018soft}). More specifically, after a certain number of NPG iterations, we reduce the temperature $\beta_S^{-1}$, and in the final stage of entropy adjustment (cf. line 10-12 in Algorithm~\ref{alg:inner}), we discard the entropy term. In this way, the policy is less susceptible to highly suboptimal local maxima in the beginning of the inner loop optimization thanks to the entropy regularization. At the end of the optimization, the variance $\sigma_j$ decreases, since the temperature is reduced and the algorithm is able to achieve a higher precision as the smoothed problem becomes a better approximation to the original one. As a result, many policy gradient updates can be saved compared to the original policy gradient method in \citet{yao2020policy}, and the quality of solutions is improved.

When the optimization by the inner loop solver is completed, the parameters $\{\mu_j\}_{j=1}^q$ are used to evaluate the reward to be back-propagated through the MC tree. More specifically, the gate sequence $\tau$ with duration $\left\{\frac{T\mathfrak{g}(\mu_j)}{ \sum_i \mathfrak{g}(\mu_i)}\right\}_{j=1}^q$ is applied and a reward is obtained. In order to deal with noisy rewards, the evaluation is repeated $m$ times, and the average reward is sent to the discrete solver. The details of the inner loop algorithm is summarized in Algorithm~\ref{alg:inner}.

\begin{algorithm}[!t]
  \caption{Improved policy gradient solver}
  \label{alg:inner}
  \begin{algorithmic}[1]
    \REQUIRE Action sequence $\bt$, number of restarts $R$, batch size $M$, learning rates $\eta_t$, total number of iterations $K$, the number of evaluation repeats $m$, the total gate duration $T$, the initial temperature $\beta_S^{-1}$, the rate of temperature decrease $0<\gamma_T<1$.\\
    \STATE Randomly initialize the mean $\{\mu_{j}\}_{j=1}^q$ and variance $\{\sigma_{j}\}_{j=1}^q$.
    \FOR {$t=1,..,R \times K$}
    \STATE Sample a batch of variables $\{\tilde\alpha_j^l\}_{j=1}^q, l\!=\!1,2, \cdots, M$ of size $M$ from sigmoid Gaussian distributions $\mathcal{SN}(\mu_{j}, \sigma_{j})$.
    \STATE Normalize the generalized QAOA parameter $\alpha_j = T\tilde\alpha_j / \sum_i \tilde\alpha_i$.
    \STATE Compute the approximate NPG direction using Eq.~\eqref{eq:NPGupdate}.
    \STATE Update the parameters with the gradient and learning rate $\eta_t$.
    \IfThen{$t\ \text{mod} \ K = 0$ and $t<(R-1)K$}{$\beta_S^{-1}\leftarrow\gamma_T\beta_S^{-1}$.}
    \IfThen{$t=(R-1)K$}{$\beta_S^{-1}\leftarrow0$.}
    \ENDFOR
    \STATE Apply the circuit $m$ times with gate sequence $\tau$ and durations $\left\{\frac{T\mathfrak{g}(\mu_j)}{ \sum_i \mathfrak{g}(\mu_i)}\right\}_{j=1}^q$, collect the rewards $\{r_k\}_{k=1}^m$, and estimate the reward $\hat{r}$ by $\hat{r}=\frac{1}{m}\sum_{k=1}^mr_k$.
    \ENSURE The mean and variance parameters $\{\mu_{j}\}_{j=1}^q$ and $\{\sigma_{j}\}_{j=1}^q$; the estimated reward $\hat{r}$.
  \end{algorithmic}
  \vspace{0.5em}
\end{algorithm}

\subsection{Relation to previous algorithms used to optimize the generalized QAOA ansatz}
We finish this section by a comparison of \agentname{} with previous methods solving the QAOA problem. As shown in Table~\ref{table:comparison}, the CD-QAOA method adopts Scipy solver for the continuous optimization, which cannot be applied to problems with noise, and the RL-QAOA method can produce suboptimal solutions in certain regimes, which we verify with numerical experiments in the next section. Moreover, note that, due to the large neural network used in RL-QAOA, it is infeasible to apply the natural gradient methods as in Section~\ref{sec:inner}.
\begin{table*}[t!]
  \centering
  \midsepremove

  \begin{tabular}{c||c|c|>{\columncolor[gray]{0.95}}c}
    \toprule
    Method                    & CD-QAOA    & RL-QAOA                     & \cellcolor[gray]{0.9}\agentname \\
    \midrule \midrule
    optimization (discrete)   & AutoReg+PG & \multirow{2}{*}{AutoReg+PG} & MCTS                            \\
    \cline{1-2}\cline{4-4}
    optimization (continuous) & SciPy      &                             & PG                              \\
    \midrule
    performance without noise & \cmark     & \cxmark                     & \cmark                          \\
    \midrule
    performance with noise    & \xmark     & \cxmark                     & \cmark                          \\
    \bottomrule
  \end{tabular}

  \midsepdefault
  \caption{Comparison among the three algorithms for the generalized QAOA ansatz: CD-QAOA, RL-QAOA and  \agentname. In this table, AutoReg+PG stands for the policy gradient algorithm with the autoregressive neural network as a policy~\citep{yao2020noise};
    \cxmark means the algorithm can fail in certain challenging regimes (e.g., large total duration $T$). }
  \label{table:comparison}
\end{table*}

\section{Numerical experiments}
\label{sec:experiments}

To benchmark the performance of \agentname{}, we consider three physics models: the $1$-dimensional Ising model, the $2$-dimensional Ising model on a square lattice, and the Lipkin-Meshkov-Glick (LMG) model. The description of the models and the additional Hamiltonians inspired from the counter-diabatic theory can be found in Appendix~\ref{sec:model}. In addition, in order to test the noise-resilience of \agentname{}, we consider three types of noise models: classical measurement Gaussian noise, quantum measurement noise, and gate rotation error, cf.~Appendix~\ref{sec:noise}.

We compare the performance of \agentname{} with that of RL-QAOA, and provide an analysis on why RL-QAOA might fail in certain regimes. Further analysis of the energy landscape of the discrete optimization reveals a surprising phenomenon: for generalized QAOA with optimal choices of the continuous degrees of freedom, there can be a large number of discrete protocols producing relatively accurate energies.

\subsection{Comparison with RL-QAOA}
\label{sec:compare_with_RL}
For the methods solving the generalized QAOA problem summarized in Table~\ref{table:comparison}, the CD-QAOA algorithm cannot be applied to problems with noise since the continuous solver is not noise-resilient, while the RL-QAOA algorithm has been shown to be effective with relatively short total duration $JT$ (using unnormalized Hamiltonians~\citet{yao2020noise}). Therefore, we use RL-QAOA as a baseline when evaluating the performance of \agentname{}, and we focus on the more challenging regime of large $JT$ with normalized Hamiltonians\footnote{The Hamiltonians used in this work are normalized by their operator norm $\|H\|$, i.e., we use $H/\|H\|$ instead of the original Hamiltonian $H$. The reason for introducing the normalized Hamiltonian is that the dependence of the cost of performing a Hamiltonian evolution $e^{-i H \alpha}$ on a quantum device -- $\Omega(\norm{H} \alpha)$ --  scales with the norm~\citet{BerryAhokasCleveEtAl2007,LowChuang2017}. Interested readers can refer to Appendix~\ref{sec:model} for more details.}.

We first compare the performance of \agentname{} against that of RL-QAOA for the physical systems discussed in Appendix~\ref{sec:model} in the presence of quantum noise. Detailed numerical results for the noiseless experiments and other noise models can be found in Appendix~\ref{sec:noise}.
In order to compare the performance of different optimizers, noisy rewards are offered to the optimizers during the training process,
and the exact rewards are only used in evaluating the protocols found by the optimizers. For \agentname{}, the protocol evaluated is given by a greedy search,~i.e., a searching process with the exploration coefficient $c=0$ in Eq.~\eqref{ucb1_update}.

\begin{figure}[t!]

  \includegraphics[width=1.0\textwidth]{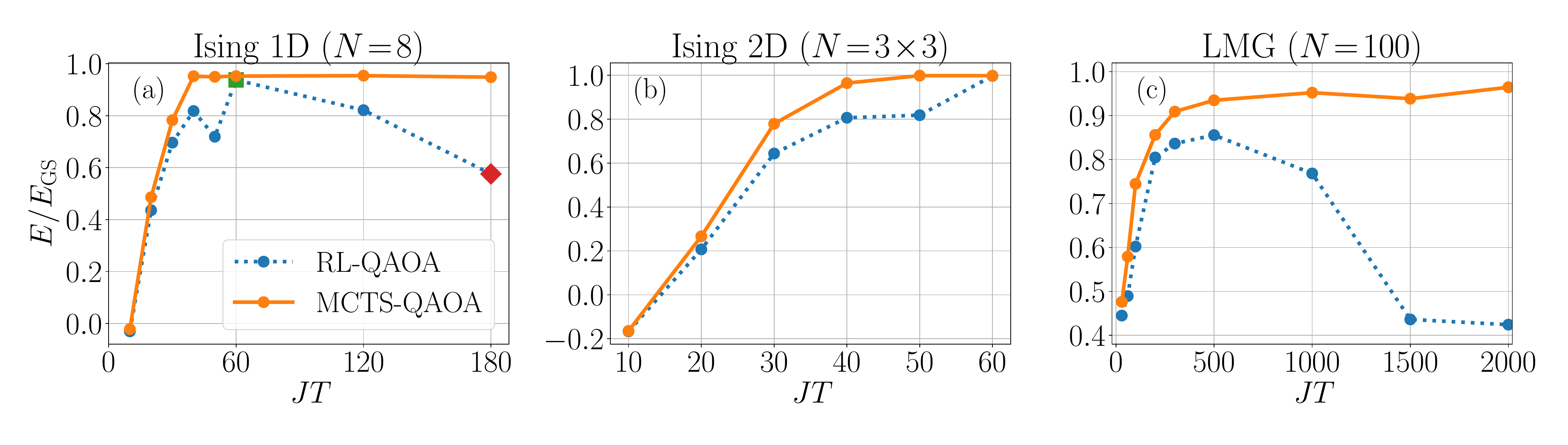}

  \vspace{-2em}
  \caption{(\textbf{Quantum noise experiment}) comparison between \agentname{} and RL-QAOA with quantum measurement noise (Appendix~\ref{sec:noise}). (a): 1D spin-$1/2$ Ising chain ($N\!=\!8$); (b): 2D spin-$1/2$ Ising chain ($N\!=\!3\times3$); (c): LMG model ($N\!=\!100$) at $h/J=0.9$.
    The blue dotted line and the orange solid line display the energy ratio $E/E_\mathrm{GS}$ obtained by RL-QAOA and \agentname{}. The green square shape and the red diamond shape in the left panel approximately corresponds to $JT\!=\!10$ (an example in the small $T$ regime) and $JT\!=\!28$ (an example in the large $T$ regime) with unnormalized Hamiltonians, respectively. The horizontal axis represents the total duration $JT$. \agentname{} outperforms RL-QAOA in all tests.}
  \label{fig:qmcts_quantum}
\end{figure}

Figure~\ref{fig:qmcts_quantum} shows the energy ratio evaluated for the protocols obtained by the optimizers across different lengths of total duration $JT$. For all three physics models, we find that the performance of \agentname{} is at least as good as that of RL-QAOA for all protocol durations.
In particular, for the 1D Ising model, \agentname{} gives protocols that find close approximations to the true ground state when $JT\gtrsim 40$;
RL-QAOA gives inferior solutions in these settings. For the 2D Ising model, while the performance of RL-QAOA is similar with that of \agentname{} at $JT=60$, the performance of RLQAOA at $JT=30, 40$ and $50$ is still inferior to that of \agentname{}. For the LMG model, the quality of the gate sequence found by RL-QAOA further decreases  when $JT>500$, and \agentname{} is significantly more robust.

The inferior performance of RL-QAOA is directly related to the joint parameterization used in RL-QAOA for the continuous and discrete policies. Since RL-QAOA optimizes the continuous and discrete variables simultaneously, for each discrete sequence, the level of accuracy of the continuous optimization can be relatively low. Consequently, the optimizer can get stuck at a suboptimal discrete sequence.

\begin{figure}[t!]

  \includegraphics[width=1.0\textwidth]{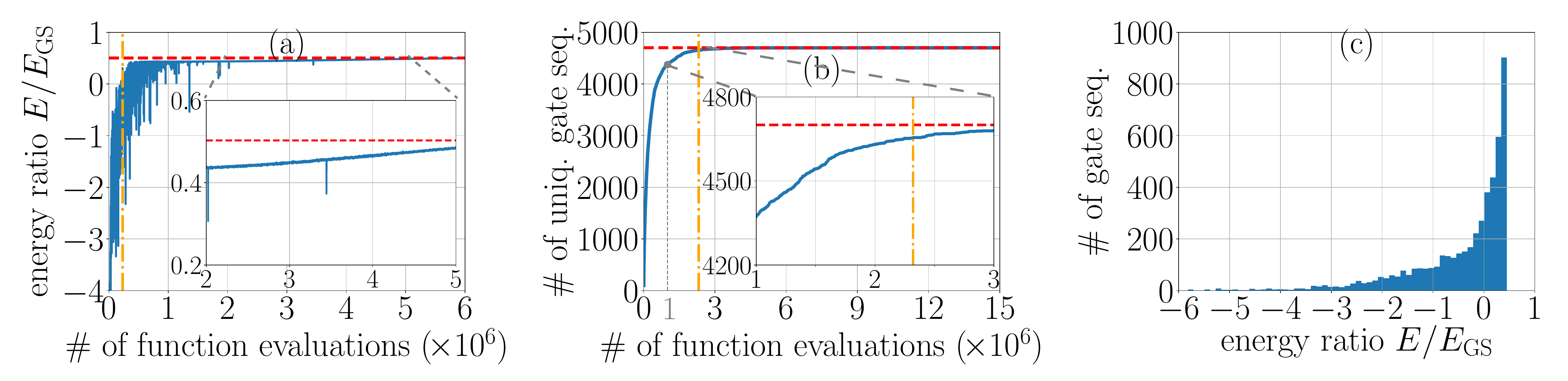}

  \caption{\label{fig:rlqaoa}\textbf{Analysis of RL-QAOA using the LMG test}: (a): Energy ratio versus number of function evaluations; (b): Number of unique gate sequences encountered versus number of function evaluations; (c): Histogram of the rewards received by the algorithm in the first 5000 iterations. The horizontal red line in the left~/~middle panel represents the maximal energy ratio and the maximal number of unique gate sequences encountered during the optimization, respectively. The orange line marks the transition between two stages of the training process.}
\end{figure}

To illustrate this behavior, we analyze the training of RL-QAOA using the LMG model with $(JT, N, q) = (1500, 100, 8)$ and noiseless rewards. Figure~\ref{fig:rlqaoa} summarizes the performance of RL-QAOA. Here by function evaluation we mean the computation of the objective function in \eqref{eq:loss}. From Figure~\ref{fig:rlqaoa}(a) and Figure~\ref{fig:rlqaoa}(b), it is clear that the training process can be divided into two distinct stages, and the transition between the two stages is marked by the dashed-dotted vertical lines\footnote{These vertical lines are drawn at the point where the number of discrete protocol gate sequences drops to $10\%$ of the total number within a single mini-batch}. In stage I, which is to the left of the vertical lines, the number of unique gate sequences encountered by RL-QAOA quickly increases, while the energy ratio keeps oscillating below zero, which suggests that RL-QAOA focuses on exploration and the continuous optimization is done only very roughly within stage I. In stage II, which is to the right of the vertical lines, the number of unique gate sequences encountered by RL-QAOA stops to grow, while the energy ratio obtained grow above zero and eventually gets stuck at around $0.5$, which means that the algorithm stops its exploration and focuses on the optimization of the continuous variables for a fixed gate sequence with stage II. The overall performance of RL-QAOA can highly depend on the discrete gate sequence that the RL-QAOA agent decides to exploit.
In the next section, we demonstrate that both the exploration and the exploitation phases in RL-QAOA can be suboptimal in this example, but the main issue is related to the suboptimal discrete sequences found in the exploration phase.

\subsection{Landscape of the discrete optimization and comparison with random search}
\label{sec:landscape}

In order to further understand the relative importance of continuous optimization versus discrete optimization for the generalized QAOA, we study the energy landscape of discrete optimization. For each discrete gate sequence, we perform numerical optimization to identify the \textit{best} continuous parameters $\{\alpha_j\}$, and record the corresponding energy ratio.

\textbf{Energy landscape of discrete optimization.} --
A profile of the discrete optimization landscape can be given by solving the corresponding continuous optimization individually on a random subset of all possible gate sequences; if the total number of possible gate sequences $\abs{\mathcal A}\qty( \abs{\mathcal A} - 1)^{q-1}$ is relatively small, this subset can actually be chosen to include all sequences. In our numerical implementation, each discrete gate sequence is sent to the natural policy gradient solver described in Section~\ref{sec:inner}, and the continuous variables are solved for different $JT$ regime. Histograms for the energy ratios obtained can then be drawn.

\begin{figure}[t!]

  \includegraphics[width=1.0\textwidth]{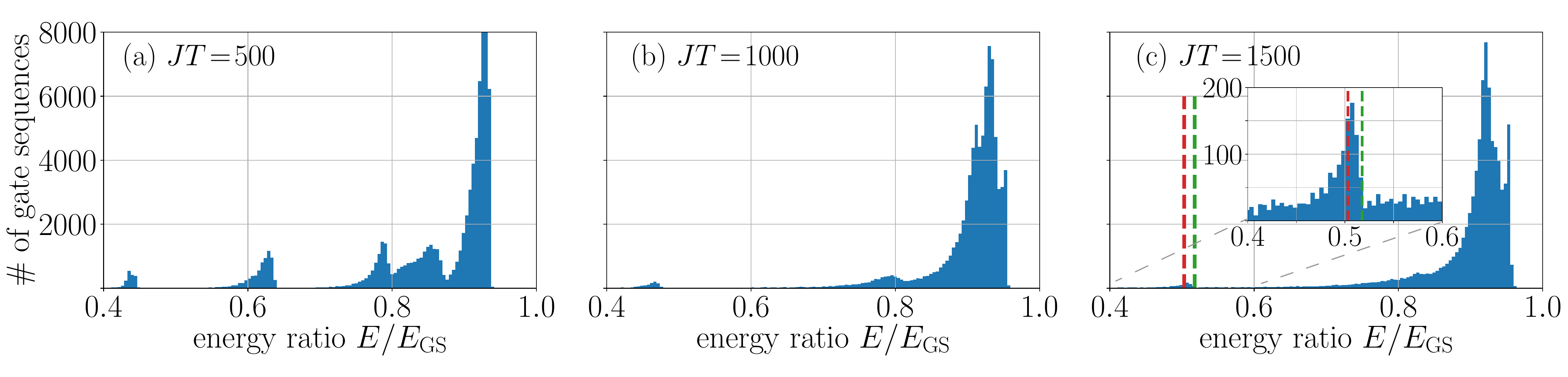}

  \includegraphics[width=1.0\textwidth]{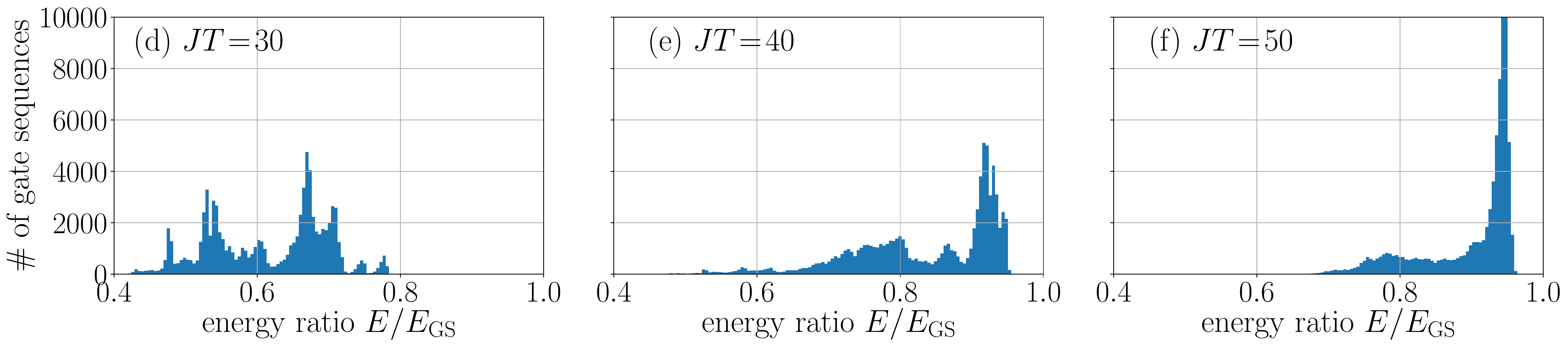}

  \caption{\textbf{Discrete landscape of the LMG model (a, b, c) and the 1D Ising model (d, e, f)}: Histograms of the energy ratio optimized by the improved natural gradient solver for $JT=500, 1000, 1500$, respectively. $N_{\text{hist}}=81920$ samples are chosen from the discrete gate sequences of generalized QAOA with parameters $q=8$, $\abs{\mathcal A}=5$ and $N=100$ (LMG) or $N=8$ (1D Ising). The dashed red line in the top right panel shows the energy ratio achieved by RL-QAOA in Figure~\ref{fig:rlqaoa}; the green dashed line shows the energy ratio obtained by the NPG solver for the same gate sequences. }
  \label{fig:landscape_LMG}
\end{figure}

Figure~\ref{fig:landscape_LMG} shows the discrete landscape for the LMG model and the 1D Ising model, respectively, where the parameters of the ansatz are $(\abs{\mathcal A}, q) \!=\! (5, 8)$, and the total number of gate sequences is thus $81920$. From the histogram plot, most gate sequences are concentrated at the right-most peak in the large $JT$ regime. Far from searching ``a needle in a haystack'', this showcases that there are plenty of ``good''\footnote{``Good'' gate sequences here means the optimized energy ratio is close to the optimal energy ratio obtainable within the generalized QAOA ansatz.} gate sequences assuming that each continuous optimization parameter is well solved.
Note that the behavior is significantly different from the discrete-only optimization, where the landscape has been shown to feature transitions between glassy, correlated and uncorrelated phases~\citep{day2019glassy}.
To the best of our knowledge, the existence of many good discrete gate sequences in the QAOA-type variational quantum algorithms has not been reported in the literature.

For the LMG model with total gate duration $JT\!=\!1500$ (cf.~Figure~\ref{fig:landscape_LMG}), while most energy ratios fall into the cluster above $0.9$, there is a smaller cluster located at $0.5$. The energy ratio obtained by RL-QAOA (cf.~Figure~\ref{fig:rlqaoa}) falls into this cluster, which is depicted by the red dashed line, while the green dashed line shows the energy ratio obtained by the natural policy gradient solver with the same gate sequence. The green line corresponds to a higher energy ratio than the red line, which means that the optimization of the continuous variables in the second stage of RL-QAOA is not as good as the NPG solver, and the difference between the two lines indicates the suboptimality caused by the exploitation. However, the suboptimality of the RLQAOA solution is mainly due to the exploration stage, since the discrete sequence that RL-QAOA chooses to exploit represents a suboptimal local optimum that belongs to a cluster much smaller than the rightmost one in the histogram. The top right panel of Figure~\ref{fig:landscape_LMG} also verifies the claim that RL-QAOA only does a rough optimization on the continuous variables before it stops exploration, since the energy ratios displayed there are mostly above $0.4$, while the energy ratio obtained in optimization stage I is mainly negative. While the landscape of the hybrid optimization is challenging for RL-QAOA, the proposed method \agentname{} is able to deal with it by using a noise-resilient solver for the continuous variables (NPG), and by exploring the discrete variables constantly using MCTS.

For the 1D Ising model with total gate duration $JT\!=\!30$ shown in the bottom left panel, where the rightmost cluster is not the largest. This means that in this setting, it is more difficult to find a gate sequence in the rightmost cluster when the random search is used. We examine the performance of random search and \agentname{} using this example in the next part.

\textbf{Comparison with random search.} --
\begin{figure}[t!]
  \centering
  \includegraphics[width=0.4\textwidth]{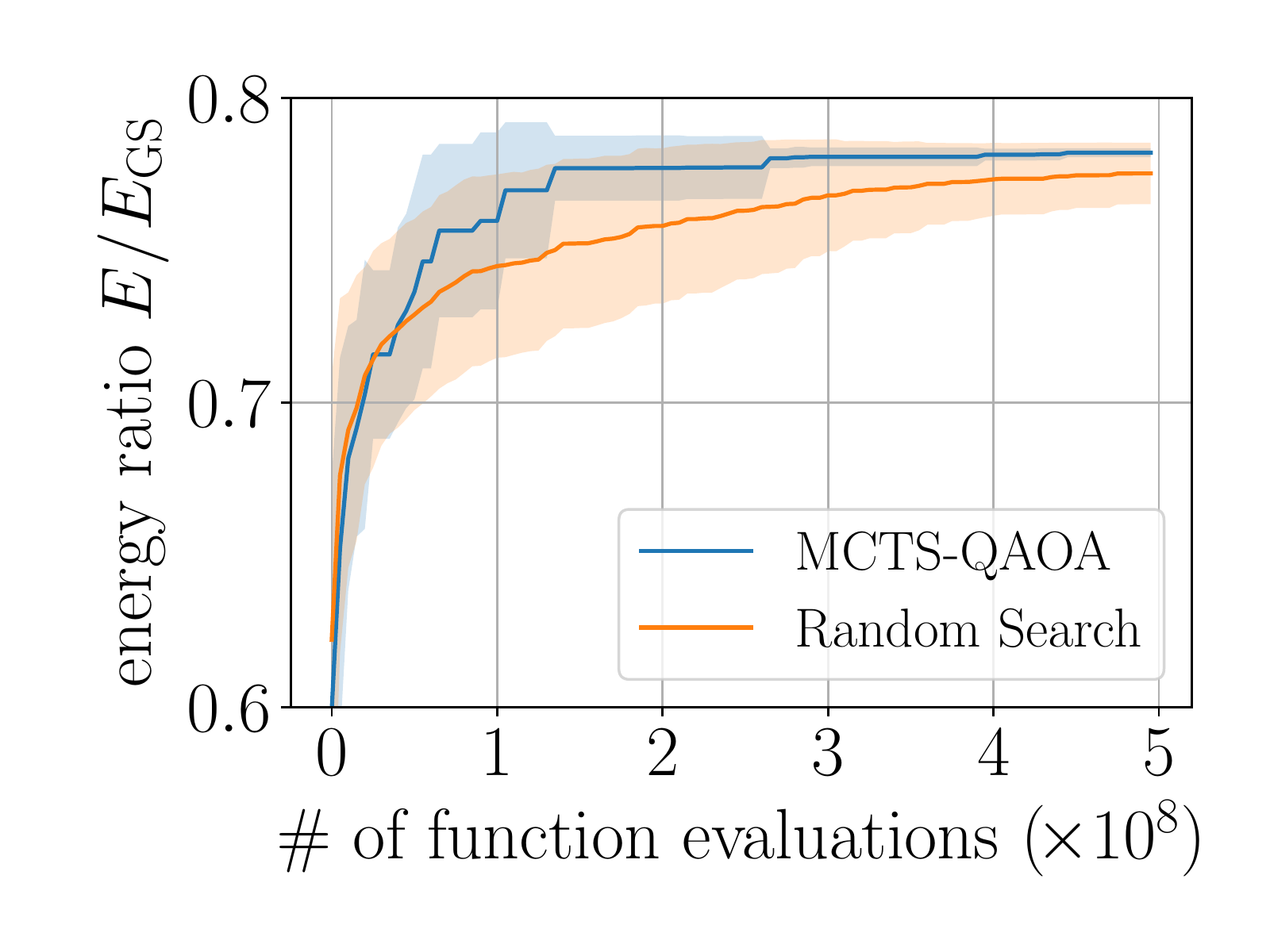}
  \caption{\textbf{Comparison between \agentname and Random Search}: The blue and orange curves correspond to \agentname{} and random search, respectively. The physics system is the 1D Ising model with duration $JT=30$, which corresponds to Figure~\ref{fig:qmcts_quantum} (a). The generalized QAOA parameters are $q=8$ and $\abs{\mathcal A}=5$. The horizontal axis is the number of function evaluations (with the function evaluation in the continuous optimization taken into account), and the vertical axis is the energy ratio. The shaded area for both algorithms represents the standard deviation across ten different random initializations.}
  \label{fig:randomsearch}
\end{figure}
A recent work \citet{mania2018simple} points out that advanced RL methods \textit{need not} outperform simpler methods such as random search. In fact, if there is no specific structure in a problem, a random search algorithm might be as efficient as any sophisticated algorithm. In addition, from the landscape illustrated in the previous histograms, one can see that, \textit{for the models we investigated}, there are lots of gate sequences with relatively high energy ratios \textit{provided that} the continuous protocols are optimized. Therefore, it is natural to compare \agentname against the random search algorithm\footnote{The random search algorithm also uses a two-level optimization, where the continuous optimization is solved by the policy gradient algorithm and the discrete optimization uses the random search. Since we assume no prior knowledge, the random search would be uniformly random on the discrete search space.}. For a fair comparison, we assume that the continuous optimization in both cases is solved by the natural policy gradient algorithm, and the difference only lies in the discrete optimization. In Figure~\ref{fig:randomsearch}, the best energy ratio in the training history is shown for the two methods, and one sees that \agentname consistently outperforms the random search across different random seeds. \agentname not only finds better gate sequences much faster, but also gives a smaller variance across different realizations. It is clear that instead of doing the search uniformly and treating each protocol as equally important, the tree statistics in \agentname better guides into a more promising search direction.

\section{Conclusion and discussions}

In this paper, we study a continuous-discrete variational quantum algorithm for the generalized QAOA ansatz. To solve this hybrid optimization problem, we design a novel algorithm that combines the Monte Carlo tree search (MCTS) algorithm, a powerful method in exploring the discrete sequence, with an improved noise-robust policy gradient solver for the continuous duration variables of a fixed gate sequence. The proposed algorithms effectively generate robust quantum control where the prior methods fail.

In this context, we expect that random search algorithms cannot efficiently determine the best gate sequence if noisy rewards are used, while \agentname is able to mitigate the noise and provide robust choice of gate sequence with the help of the tree structure it maintains.
Moreover, it is possible for \agentname{} to further reduce the number of evaluations by assigning different number of iterations for different gate sequences, e.g., it can assign more iterations for the more promising gate sequences.
Also, \agentname{} allows for the application of transfer learning using the tree statistics, which is not possible for the random search.

There are a number of possible ways to extend the problem presented in this paper:

\noindent\textbf{Learning based guided search.} --
MCTS can be possibly guided by a learned functional approximator, such as neural networks or tensor networks. We have also tried the implementation of AlphaZero in the same experimental settings. However, the neural network based method does not work better than the simple MCTS. We find that the value function mapping from the discrete gate sequences to the score was quite hard to learn. One reason might be that the continuous policy gradient will try the best to optimize the energy ratio to the highest, thus making this mapping from discrete sequences to score, highly non-linear. Also, in terms of sampling efficiency, the neural network based approach needs lots of samples to fit the function, which is a heavy overhead compared to the simple MCTS approach. Nevertheless, the question remains open as to how to upgrade MCTS to a guided search.

\noindent\textbf{Amortized computation.} --
The computation within the policy gradient solver for different gate sequences can possibly be amortized. Currently, the continuous and discrete optimizations are separated. If some functional can be learned by replaying the data during the policy gradient iteration, the number of function evaluations can be further reduced. However, one difficulty in the quantum setting is that we do not have access to the quantum state, and thus we cannot learn a mapping taking the quantum state as input, unless we apply non-trivial quantum tomography. Therefore, how to reuse the past information and make the \agentname{} algorithm quickly adaptive in physical setting remains to be investigated. Advanced algorithms like meta learning can be explored in the future work.

\noindent\textbf{Budget-aware variational quantum algorithms.} -- A point of high interest is the design of budget-aware variational quantum algorithms. The importance of sample efficiency in the quantum setting can never be overemphasized. Each run of a quantum circuit can be expensive and quantum decoherence noise is usually not stationary over time. The budget-awareness property can be naturally incorporated in the MTCS framework. Making use of the tree structure, the adaptive algorithm would distribute more function evaluation budget to the most-visited or more promising nodes. The current algorithm likely operates in a budget-sufficient regime and uses the same amount of budget for each discrete gate sequences. We hope the adaptive algorithm can hit the sweet spot in the middle, i.e., use the right amount of computational budget and still compute the best possible gate sequence design. We hope that the present work will accelerate the research of budget-aware variational quantum algorithms in a realistic setting.

\section*{Acknowledgement}

We thank Michael Luo for donating eight NVIDIA V100 Tensor Core GPUs to support computation. This work was partially supported by the Department of Energy under Grant No. DE-AC02-05CH11231 and No. DE-SC0017867 (L.L., J.Y.), and by the National Science Foundation under the NSF QLCI program through grant number OMA-2016245 (L.L.). M.B.~was supported by the Marie Sklodowska-Curie grant agreement No 890711.

We used  W\&B~\cite{wandb} to organize and analyze the experiments. The reinforcement learning networks are implemented in NumPy~\cite{harris2020array}, and Jax~\cite{jax2018github}; the quantum systems are simulated in \href{https://github.com/weinbe58/QuSpin\#quspin}{Quspin}~\cite{weinberg2017quspin, weinberg2019quspin}. We thank Berkeley Research Computing (BRC) and Google Cloud Computing Services (GCP) for providing the computational resources.

\newpage
\appendix

\clearpage

\section{Related works}
\label{app:related}

\noindent\textbf{Hybrid optimization}: The generalized QAOA ansatz introduces a discrete and continuous control problem: the discrete degrees of freedom are the gates/unitaries that define the control protocol, while the continuous degrees of freedom are the gate duration. Most reinforcement learning algorithms~\citep{lillicrap2015continuous, Bertsekas2019ReinforcementLA, trabucco2021conservative} typically deal with the control of either discrete or continuous degree of freedom, and hardly consider the discrete and continuous control simultaneously in the policy. Even though the continuous control can always be discretized, it is always beneficial and desirable to consider discrete and continuous variables together, without loss of the flexibility of continuous control. Furthermore, the idea of continuous and discrete optimization can be quite general, and shows up in real world application like robotics~\citep{neunert2020continuous, delalleau2019discrete} and strategic games~\citep{vinyals2019grandmaster}. Combining the discrete and continuous control together, the control capability of the algorithm can be quickly enhanced. In general, the discrete variables are usually chosen as the categories of actions, and the continuous variables will are naturally given by the strength for each specific action. Our work aims to shed light on the hybrid control in the field of quantum control, and we also hope it will accelerate the research of hybrid discrete-continuous optimization algorithms in the wider community.

\noindent\textbf{Counter diabatic driving}: Counter diabatic driving~\citep{sels2017minimizing, hegade2022digitized}, an example of a shortcut to adiabaticity (STA), introduces an extra auxiliary counter-diabatic (CD) Hamiltonian to suppress transitions (or excitations) between instantaneous eigenvalues.

For a given quantum state $\left|\psi\right\rangle$ evolving under a time dependent Hamiltonian $H_0(\lambda(t))$, the Schr\"odinger equation reads as
\begin{eqnarray}
  & i \hbar\partial_t \left|\psi\right\rangle= H_0(\lambda(t)) \left|\psi\right\rangle, \nonumber\\ &\left|\psi_{i}\right\rangle =\left|\psi_{\mathrm{GS}}(\lambda =0)\right\rangle, \left|\psi_{*}\right\rangle =\left|\psi_{\mathrm{GS}}(\lambda =1)\right\rangle.
\end{eqnarray}
In the rotating frame, Hamiltonian remains stationary under the unitary transformation $U(\lambda(t))$,~i.e.~in the instantaneous eigenbasis of Hamiltonian $H_0(\lambda)$. The wave function $|\tilde \psi \rangle=U(\lambda)|\psi \rangle$ in the rotating frame satisfies the following Schr\"odinger equation:
\begin{equation}
  i \hbar\partial_t |\tilde\psi\rangle=\left( \tilde {H_0}(\lambda(t)) -\dot{\lambda}\tilde{\mathcal A}_\lambda \right)|\tilde \psi\rangle,
\end{equation}
where $\tilde {H_0}(\lambda(t))=U^\dagger H_0(\lambda(t)) U,\tilde {\mathcal A}_\lambda=i U^\dagger \partial_\lambda U$. Specifically, instead of being diagonalized, the original Hamiltonian picks up an extra contribution due to the change in the parameter $\lambda(t)$, and the effective Hamiltonian becomes
\begin{equation}
  H^{\rm eff}_0=\tilde {H_0}-\dot{\lambda} \tilde{\mathcal A}_\lambda.
\end{equation}
The idea of the CD driving is to evolve the system with the counterdiabatic Hamiltonian
\begin{equation}
  H_{\rm CD}(t)=H_0{+\dot\lambda \mathcal A_\lambda}.
\end{equation}
Importantly, in the moving frame $H_{\rm CD}^{\rm eff}(t)=\tilde {H_0}$ is \textit{stationary and no transitions occur}.

However, in practice, the precise counter-diabatic Hamiltonian is intractable and usually approximated by different methods. A good number of prior works~\citep{PhysRevResearch.2.013283, hartmann2019rapid, hegade2022digitized, hegade2021portfolio, zhou2020experimental, wurtz2021counterdiabaticity, hegade2020shortcuts} are based on the concept of a variational approximation to the CD Hamiltonian~\citep{sels2017minimizing}. Most of these works typically make use of an analytically computed expression available for few-qubit systems; they first derive the continuous form of the variational gauge potential, and then discretize the underlying dynamics using the Trotter-Suzuki formula
In this work, we aim to bypass these constraints by applying the variational generalized QAOA ansatz using additional gates, generated by terms that occur in the approximation to the  variational adiabatic gauge potential. These extra gates can provide a shortcut to the preparation of the ground state, compared to the original alternating QAOA ansatz. Physically, this shortcut results in shorter circuit simulation times, which provices a significant advantage on noisy NISQ devices.

\noindent\textbf{AutoML and neural architecture search}: Automatic machine learning or AutoML has recently attracted lots of attentions as it reduces human efforts in designing the neural architecture from experience and instead leverage the computational power to search the best configuration. One of the most pronounced examples are neural architecture search (NAS) and their variants~\citep{zoph2016neural, elsken2018neural, liu2018darts, cai2018proxylessnas, real2020automl}, where reinforcement learning or evolutionary strategies are used to find a better network architecture. Inspired by the success of AutoML, the architecture of quantum circuits can also be improved by machine learning algorithms, such as the quantum version of Neural Architecture Search~\citep{wang2021quantumnas, wang2021roqnn, zhang2021neural, zhang2020differentiable, kuo2021quantum}. These prior works interpret the problem as quantum compiling problems, which assembles quantum gates in the low level. Instead of exposing a huge number of choice alternatives for the search algorithms, our work specially uses the variational gauge potentials as the Hamiltonian pool for the search algorithm in a computation-efficient way. Compared with QAOA, \agentname{} has more degree of freedom to approximate the unitary operator; compared with the quantum compiling, it does not search gates in the low level due to the constraint of computations. From this perspective, our method hits the sweet spot between the expressivity and efficiency.

\section{Setup of physical models}
\label{sec:model}

We first give a brief review on the physical models used in the numerical experiments.
In all experiments, we choose the target state as the ground state of the Hamiltonian $H$, denoted $\ket{\pgs(H)}$. The spin-$1/2$ matrices describing spin $i$ are denoted by $X_i, Y_i, Z_i$. In contrast to the models considered in \citet{yao2020noise,yao2020reinforcement}, the Hamiltonians used in this work is normalized by its operator norm $\|H\|$, i.e., we use $H/\|H\|$ instead of the original Hamiltonian $H$. The reason for introducing the normalized Hamiltonian is as follows. For generic Hamiltonians $H$ (e.g., sparse matrices),  the cost of performing a Hamiltonian evolution $e^{-i H \alpha}$ on a quantum device is $\Omega(\norm{H} \alpha)$~\citet{BerryAhokasCleveEtAl2007,LowChuang2017}. Due to the potential differences between the Hamiltonian norms in the Hamiltonian pool $\mathcal{A}$, using a normalized Hamiltonian $H/\norm{H}$ (the corresponding duration parameter $\alpha$ is thus multiplied by $\norm{H}$) can lead to a more realistic estimate of the cost of the quantum simulation.  Due to this multiplication factor, the duration shown in the results below is larger than that presented in~\citet{yao2020noise,yao2020reinforcement}.

\noindent\textbf{One-dimensional (1D) Ising model}

The spin-$1/2$ Ising Hamiltonian reads as:
\begin{eqnarray}
  H \!&=&\!H_1\!+\!H_2, \quad
  H_1\!\!=\!\! \sum_{i=1}^N J Z_{i+1}Z_i \!+\! h_z Z_i,\quad
  H_2= \sum_{i=1}^N h_x X_i, \nonumber
  \label{eqn:ising}
\end{eqnarray}
where $N$ is the number of qubits and the parameters are set as $h_z/J\!=\!0.4523$ and $h_x/J\!=\!0.4045$ \citep{kim2013ballistic}. These parameters are close to the critical line of the model in the thermodynamic limit, where the quantum phase transition occurs. They are also reported in Ref.~\citep{matos2020quantifying} to be in the most challenging parameter region using QAOA. We use periodic boundary conditions here. The initial state for this experiment is given by $z$-polarized product state,~i.e.~$|\psii\rangle\!=\!|\!\!\uparrow\cdots\uparrow\rangle$.

For the Hamiltonian pool, we use $\mathcal{A}\!=\!\left\{
  J\frac{H_1}{||H_1||},
  J\frac{H_2}{||H_2||},
  J\frac{A_1}{||A_1||},
  J\frac{A_2}{||A_2||},
  J\frac{A_3}{||A_3||}
  \right\},$ where $A_1 \!=\! \sum_{i=1}^N  Y_{i},  \
  A_2 \!=\! \sum_{i=1}^N  X_{i}Y_{i} + Y_{i}X_{i},\
  A_3 \!=\! \sum_{i=1}^N  Z_{i}Y_{i} + Y_{i}Z_{i}$.
The operators $A_j$ are precisely the first three terms in the expansion for the adiabatic gauge potential of the translation-invariant 1D Ising model~\citep{yao2020reinforcement}.

\noindent\textbf{Two-dimensional (2D) Ising model}

The 2D spin-$1/2$ transverse-field Ising model reads:
\begin{eqnarray}
  H \!=\!H_1\!+\!H_2,\quad
  H_1\!\!=\!\! J\sum_{\langle i,j\rangle} Z_i Z_j + h_z \sum_{j} Z_j,\;
  H_2= \sum_{j} h_x X_j, \nonumber
  \label{eqn:ising-2d}
\end{eqnarray}
where $\langle i,j\rangle$ denotes nearest neighbors on the square lattice. The model parameters are set as $h_z/J\!=\!2$ and $h_x/J\!=\!3$. The initial state is $|\psii\rangle\!=\!\ket{\uparrow}$,~i.e.~$z$-polarized product state on 2D lattice.

For the Hamiltonian pool, we use $\mathcal{A}\!=\!\left\{
  J\frac{H_1}{||H_1||},
  J\frac{H_2}{||H_2||},
  J\frac{A_1}{||A_1||},
  J\frac{A_2}{||A_2||},
  J\frac{A_3}{||A_3||}
  \right\}$, where $A_1 \!=\! \sum_{j} Y_j, \
  A_2 \!=\! \sum_{\langle i,j\rangle} X_i Y_j + Y_i X_j,\
  A_3 \!=\! \sum_{\langle i,j\rangle} Z_i Y_j + Y_i Z_j$.

\noindent\textbf{Lipkin-Meshkov-Glick (LMG) model}

The Lipkin-Meshkov-Glick (LMG) model~\citep{lipkin1965validity} reads:
\begin{eqnarray}
  H \!=\!H_1\!+\!H_2,\quad
  H_1\!\!=\!\! -\frac{J}{N}\sum_{i,j=1}^N X_i X_j,\;
  H_2=  h\sum_{j=1}^N \left(Z_j+\frac{1}{2}\right),\nonumber
  \label{eqn:LMG}
\end{eqnarray}
where $J$ is the interactions trength, and $h$ stands for the magnetic field strength. The LMG model preserves the total spin, and the ground state is contained in an $N\!+\!1$ dimensional subspace due to this symmetry. This makes the LMG model particularly interesting because it allows us to simulate its dynamics for a large number of spins, where many-body effects, such as collective phenomena, dominate the physics of the system.

For instance, in the thermodynamic limit $N\to\infty$, the LMG model exhibits a quantum phase transition at $h_c/J\!=\!1$~\citep{botet1983large}. The transition is between a ferromagnetic (FM) order in the ground state in the $x$-direction ($h/J\ll 1$), and the paramagnetic order ($h/J\gg 1$).

For the Hamiltonian pool, we use $\mathcal{A}\!=\!\left\{
  J\frac{H_1}{||H_1||},
  J\frac{H_2}{||H_2||},
  J\frac{A_1}{||A_1||},
  J\frac{A_2}{||A_2||},
  J\frac{A_3}{||A_3||}
  \right\}$,   where
\begin{eqnarray}
  A_1 &=&\sum_{j=1}^N Y_j , \nonumber \\
  A_2 &=& \frac{1}{N} \qty( \sum_{j=1}^N Y_j ) \qty( \sum_{j=1}^N X_j ) + \frac{1}{N} \qty( \sum_{j=1}^N X_j ) \qty( \sum_{j=1}^N Y_j ),\nonumber\\
  A_3 &=& \frac{1}{N} \qty( \sum_{j=1}^N Y_j ) \qty( \sum_{j=1}^N \left( Z_j+\frac{1}{2} \right) )  + \frac{1}{N} \qty( \sum_{j=1}^N \left( Z_j+\frac{1}{2} \right) ) \qty( \sum_{j=1}^N Y_j ).
\end{eqnarray}

\section{Noise models}
\label{sec:noise}

An essential part of our study is the performance of the algorithms in the presence of noise. As mentioned in the main text, noise sets the current bottle neck for reliable quantum computation. Therefore, it is of primary importance for the near-term utility of quantum computers to develop stable and noise-robust manipulation algorithms.

We use the following three noise models in our numerical experiments:
(i) classical measurement noise,
(ii) quantum measurement error which micmic the situation on present-day NISQ devices, and
(iii) gate rotation error noise.

\noindent\textit{\textbf{Classical measurement Gaussian noise}} is added to the cost function according to
$$\mathcal{L}_\gamma (\params)= \mathcal{L}(\params) + \epsilon_\gamma, $$ where $\epsilon_\gamma \sim \mathcal N (0, \gamma^2)$ and $\gamma$ denotes the noise strength, and $\mathcal N$ is the normal distribution. Gaussian noise models various kinds of uncertainty present in experiments using an additive Gaussian random variable, which follows from the Central Limit theorem.

\noindent\textit{\textbf{Quantum measurement noise}}:
$$
  \mathcal{L}_Q (\params)= \mathcal{L}(\params) + \epsilon_Q,
$$
where the noise strength depends on the strength of the energy quantum fluctuations
$$
  \Delta \mathcal{E} = N^{-1} \sqrt{ \langle \psi(T)|H^2|\psi(T)\rangle - \langle \psi(T)|H|\psi(T)\rangle^2 },
$$ and $\epsilon_Q$ is randomly sampled from $\mathcal N (0, \Delta \mathcal{E}^2)$. Quantum noise models the uncertainty arising from quantum measurements. For instance, quantum fluctuations are large when the evolved quantum state is far away from the target, while they decrease when the final state approaches the target ground state.

\noindent\textit{\textbf{Gate rotation error noise}}: $$
  \mathcal{L}_\delta (\params)= \mathcal{L}(\params'),\;\; \params'=(\{\alpha_i  + \alpha_i \epsilon_i \}_{i=1}^q, \bt)
$$
where gate error strengths are multiplicative and the corresponding ratios are $\epsilon_{i} \sim \mathcal N (0, \delta^2)$ for some simulation parameter $\delta$ which controls the noise strength. Gate rotation errors~\cite{sung2020exploration} present yet another common noise source, which arises due to imperfections or lack of calibration in the quantum computer hardware.

\section{Details for the natural policy gradient with entropy regularization}\label{app:npg}

For a general $d$-dimensional Gaussian distribution $\mathcal{N}(\mu, \Sigma)$, the Shannon entropy is defined as $\mathbb{E}(-\log(p(x)))$, where $p(x) = (2\pi)^{-\frac{d}{2}}|\Sigma|^{-\frac{1}{2}}\exp(-\frac{1}{2}(x-\mu)^\top\Sigma^{-1}(x-\mu))$. Hence
\[
  \begin{aligned}
    \mathbb{E}(-\log(p(x))) & =-\mathbb{E}\log\left[(2\pi)^{-\frac{d}{2}}|\Sigma|^{-\frac{1}{2}}\exp(-\frac{1}{2}(x-\mu)^\top\Sigma^{-1}(x-\mu))\right] \\
                            & =\mathbb{E}\left[\frac{d}{2}\log2\pi+\frac{1}{2}\log|\Sigma|+\frac{1}{2}(x-\mu)^\top\Sigma^{-1}(x-\mu)\right]             \\
                            & =\frac{d}{2}\log2\pi+\frac{1}{2}\log|\Sigma|+\frac{1}{2}\mathbb{E}(x-\mu)^\top\Sigma^{-1}(x-\mu)
    \\
                            & =\frac{d}{2}\log2\pi+\frac{1}{2}\log|\Sigma|+\frac{1}{2}\mathbb{E}\Tr((x-\mu)^\top\Sigma^{-1}(x-\mu))                     \\
                            & =\frac{d}{2}\log2\pi+\frac{1}{2}\log|\Sigma|+\frac{1}{2}\mathbb{E}\Tr(\Sigma^{-1}(x-\mu)(x-\mu)^\top)                     \\
                            & =\frac{d}{2}\log2\pi+\frac{1}{2}\log|\Sigma|+\frac{d}{2}.
  \end{aligned}
\]
Omitting the constants, it is equivalent to take the entropy as $\frac{1}{2}\log|\Sigma|$. For the model used, the probability distribution is a product of normal distribution, i.e., $\Sigma$ is a diagonal matrix with length $q$ and diagonal elements $\sigma_i$, so the corresponding entropy function is $\mathbb{E}(-\log(p(x)))=\sum_{i=1}^q\log \sigma_i$.

In the implementation, we adopt the parameterization $\sigma_i = \exp(t_i)$ to assure that $\sigma_i$ is positive. Then for the distribution $\mathcal{N}(\mu_i, \sigma_i)$, we have
\[
  \log p_i(x) = -\frac{(x-\mu_i)^2}{2\sigma_i^2}-\log\sigma_i - \frac{1}{2}\log(2\pi) = -\frac{1}{2}(x-\mu_i)^2e^{-2t_i}-t_i-\frac{1}{2}\log(2\pi),
\]
and
\[
  \nabla\log p_i(x) = ((x-\mu_i)e^{-2t_i}, (x-\mu_i)^2e^{-2t_i}-1)^\top,
\]
where the gradient is taken with respect to the parameters. Since $\{\delta_i\}_{i=1}^q$ are independent, the Fisher information matrix is a block diagonal matrix with the $i$-th block equal to
\[
  F_i=\mathbb{E}\nabla\log p_i(x)\nabla\log p_i(x)^\top =
  \mathbb{E}\begin{bmatrix}
    \frac{(x-\mu_i)^2}{\sigma_i^4}                            & \frac{(x-\mu_i)^3}{\sigma_i^3}-\frac{(x-\mu_i)}{\sigma_i}        \\
    \frac{(x-\mu_i)^3}{\sigma_i^3}-\frac{(x-\mu_i)}{\sigma_i} & \frac{(x-\mu_i)^4}{\sigma_i^4}-\frac{2(x-\mu_i)^2}{\sigma_i^2}+1
  \end{bmatrix}
  = \begin{bmatrix}
    \frac{1}{\sigma_i^2} & 0 \\
    0                    & 2
  \end{bmatrix}.
\]
Recall that for a fixed gate sequence $\bt$, we set $R(\bm\delta) = -E\left(\left\{\frac{T\mathfrak{g}(\delta_j)}{\sum_k\mathfrak{g}(\delta_k)}\right\}_{j=1}^q,\bt\right)/N$, where $\mathfrak{g}$ denotes the sigmoid function, and
\[
  \mathcal{J}(\{\mu_{j}, \sigma_{j}\}_{j=1}^q)\!= \mathbb {E}_{\substack{\delta_j\sim \mathcal {N}(\mu_{j}, \sigma_{j}) }}R(\bm \delta) + \beta^{-1}_{{S}} \sum_{j=1}^q \log\sigma_j.
\]
Hence the gradient of $\mathcal{J}$ is
\[
  \mathbb{E}R(\bm\delta)\nabla\log p(\bm\delta)+\beta^{-1}_{{S}}\nabla\sum_{j=1}^q \log\sigma_j,
\]
where the gradient is taken with respect to the parameters, and $p(\bm\delta) = \prod_{i=1}^qp_i(\delta_i)$. Therefore, the unbiased estimators for the variables are
\[
  \begin{bmatrix}
    \frac{\partial \mathcal{J}}{\partial \mu_j} \\\frac{{\partial\mathcal{J}}}{\partial t_j}
  \end{bmatrix}
  \leftarrow\begin{bmatrix}
    R(\bm \delta)\xi_j/\sigma_j \\
    R(\bm \delta)(\xi_j^2-1) + \beta^{-1}_{{S}}
  \end{bmatrix},
\]
where $\xi_j$ are independent standard normal variables and $\delta_j = \sigma_j\xi_j+\mu_j$.
As a result, the unbiased estimators for the natural gradient direction become
\[
  F_j^{-1}\begin{bmatrix}
    \frac{\partial \mathcal{J}}{\partial \mu_j} \\\frac{{\partial\mathcal{J}}}{\partial t_j}
  \end{bmatrix}\leftarrow
  \begin{bmatrix}
    \sigma_jR(\bm\delta)\xi_j \\
    \frac{1}{2}R(\bm\delta)(\xi_j^2-1) + \frac{1}{2}\beta^{-1}_{{S}}
  \end{bmatrix},
\]
since $F$ is a block diagonal matrix with the $j$-th block given by $F_j$.

\section{Additional experiment results}
In Section~\ref{sec:compare_with_RL}, we have presented a comparison between the RL-QAOA method and \agentname{} for three different physics models with the quantum noise. In this section, we report the test results with the other types of noise, namely the results with the Gaussian noise, the results with the gate rotation error, and the results when no noise is considered (cf. Appendix~\ref{sec:noise}). We can observe from the comparison that \agentname's performance is much more stable and accurate.
\begin{figure}[t!]

  \includegraphics[width=1.0\textwidth]{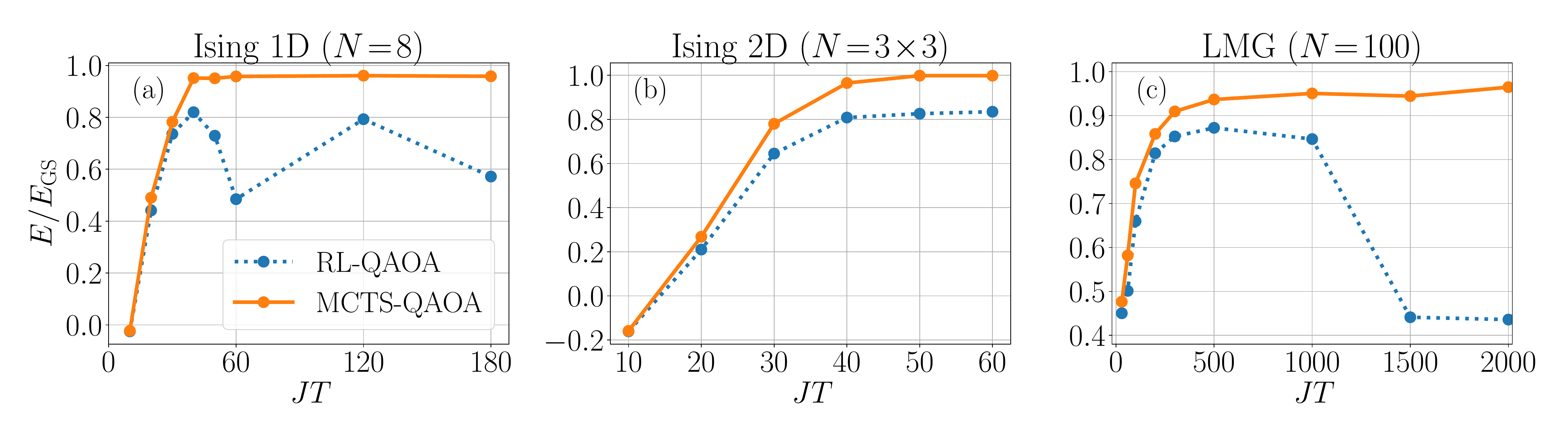}

  \includegraphics[width=1.0\textwidth]{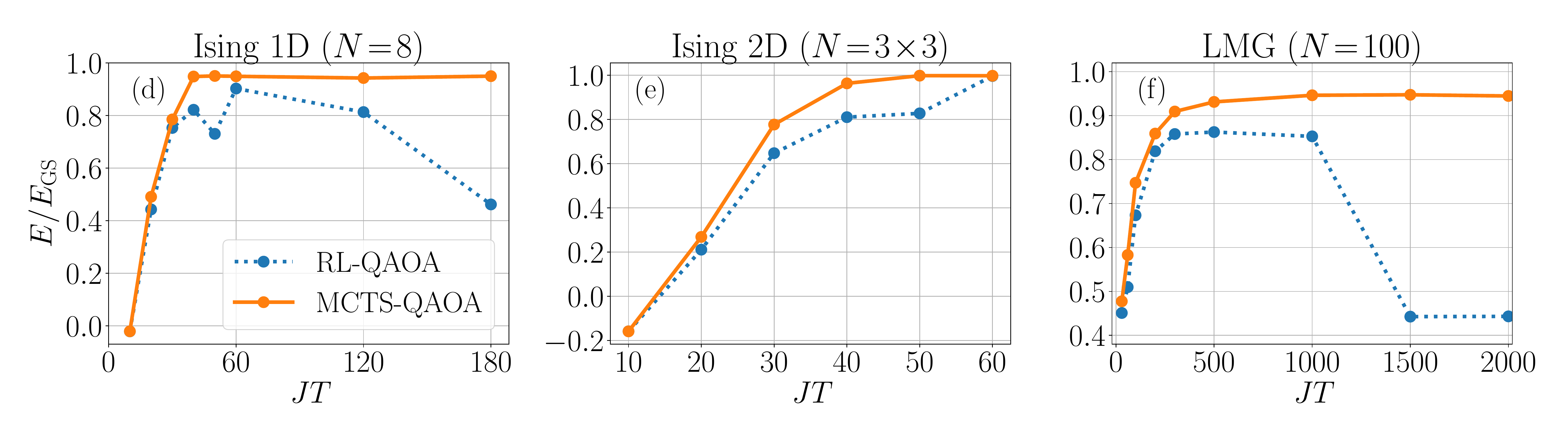}

  \includegraphics[width=1.0\textwidth]{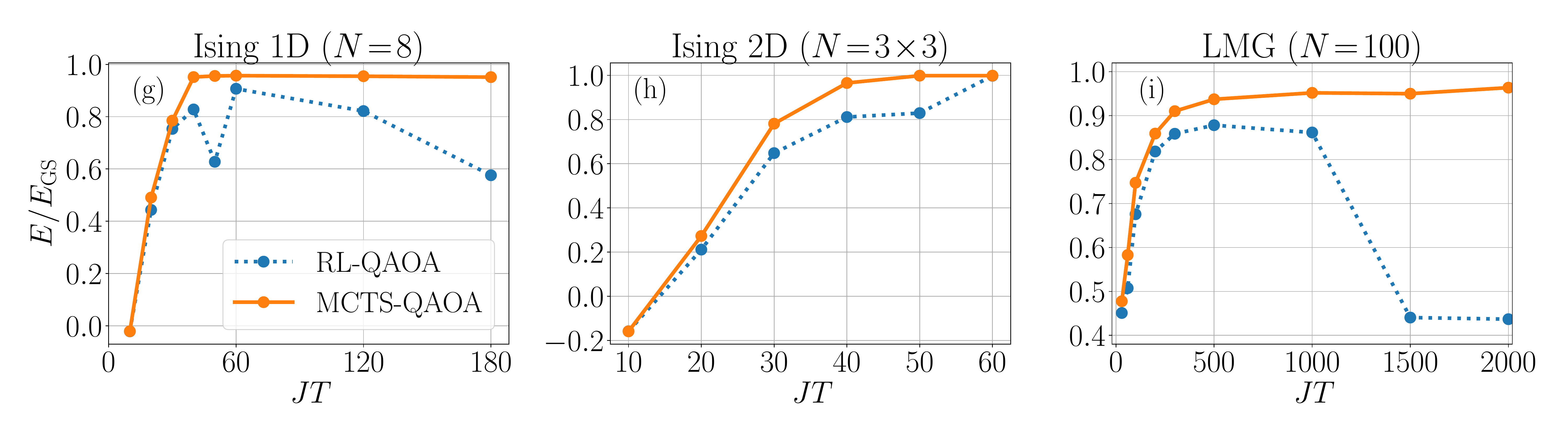}

  \vspace{-2em}
  \caption{(\textbf{experiment with other types of noise models or without noise}) comparison between \agentname{} and RL-QAOA. The physics setup is the same as that in Figure~\ref{fig:qmcts_quantum}. (a-c): Gaussian noise with $\gamma=0.1$; (d-f): gate rotation noise with $\delta=0.1$; (g-i): experiments without noise (cf. Appendix~\ref{sec:noise}). }
  \label{fig:qmcts_other_noise}
\end{figure}

From Figure~\ref{fig:qmcts_other_noise}, one can observe similar behavior the two methods as in Section~\ref{sec:compare_with_RL}, i.e., \agentname{} outperforms RL-QAOA in all settings and the gaps grow larger in the regime of large total gate durations. The raw data for the energy ratio obtained by \agentname{} is summarized in Table~\ref{tab:qmcts_fid} (highlighted in bold), which offers a more visually and quantitatively convenient comparison across different models.

\begin{table}[!htbp]
  \centering
  \begin{tabular}{l||cc||cc}
    \toprule
    \cellcolor[HTML]{D4D4D4}                                               & \cellcolor[HTML]{D4D4D4}Gate rotation noise                        & \cellcolor[HTML]{D4D4D4}Quantum noise                   & \cellcolor[HTML]{D4D4D4}Gaussian noise                  & \cellcolor[HTML]{D4D4D4}No noise                        \\
    \multicolumn{1}{c||}{\multirow{-2}{*}{ \cellcolor[HTML]{D4D4D4} $JT$}} & \multicolumn{4}{c}{\cellcolor[HTML]{D4D4D4} $(E / E_{\text{GS}})$}                                                                                                                                                                               \\
    \midrule
    \midrule
    (\textit{Model})                                                       & \multicolumn{4}{c}{\cellcolor[HTML]{E5E4E2} (a) \textbf{Ising 1D}}                                                                                                                                                                               \\
    $10.0$                                                                 & {$-0.0208$} (\textcolor{gray}{\textit{$-0.0208$}})                 & {$-0.0219$} (\textcolor{gray}{\textit{$-0.0238$}})      & {$-0.0225$} (\textcolor{gray}{\textit{$-0.0209$}})      & {$-0.0210$} (\textcolor{gray}{\textit{$-0.0207$}})      \\
    $20.0$                                                                 & {$0.4907$} (\textcolor{gray}{\textit{$0.4884$}})                   & {$0.4862$} (\textcolor{gray}{\textit{$0.4863$}})        & {$0.4903$} (\textcolor{gray}{\textit{$0.4905$}})        & {$0.4907$} (\textcolor{gray}{\textit{$0.4908$}})        \\
    $30.0$                                                                 & {$0.7849$} (\textcolor{gray}{\textit{$0.7844$}})                   & {$0.7830$} (\textcolor{gray}{\textit{$0.7796$}})        & {$0.7825$} (\textcolor{gray}{\textit{$0.7833$}})        & {$0.7850$} (\textcolor{gray}{\textit{$0.7850$}})        \\
    $40.0$                                                                 & {$0.9481$} (\textcolor{gray}{\textit{$0.9486$}})                   & {$0.9521$} (\textcolor{gray}{\textit{$0.9477$}})        & {$0.9512$} (\textcolor{gray}{\textit{$0.9513$}})        & {$0.9516$} (\textcolor{gray}{\textit{$0.9527$}})        \\
    $50.0$                                                                 & {$0.9503$} (\textcolor{gray}{\textit{$0.9499$}})                   & {$0.9499$} (\textcolor{gray}{\textit{$0.9564$}})        & {$0.9505$} (\textcolor{gray}{\textit{$0.9581$}})        & {$0.9559$} (\textcolor{gray}{\textit{$0.9574$}})        \\
    $60.0$                                                                 & {$0.9489$} (\textcolor{gray}{\textit{$0.9614$}})                   & {$0.9526$} (\textcolor{gray}{\textit{$0.9540$}})        & {$0.9576$} (\textcolor{gray}{\textit{$0.9560$}})        & {$0.9570$} (\textcolor{gray}{\textit{$0.9621$}})        \\
    $120.0$                                                                & {$0.9424$} (\textcolor{gray}{\textit{$0.9495$}})                   & {$0.9543$} (\textcolor{gray}{\textit{$0.9548$}})        & {$0.9606$} (\textcolor{gray}{\textit{$0.9524$}})        & {$0.9548$} (\textcolor{gray}{\textit{$0.9602$}})        \\
    $180.0$                                                                & {$0.9495$} (\textcolor{gray}{\textit{$0.9415$}})                   & {$0.9486$} (\textcolor{gray}{\textit{$0.9502$}})        & {$0.9582$} (\textcolor{gray}{\textit{$0.9556$}})        & {$0.9514$} (\textcolor{gray}{\textit{$0.9543$}})        \\
    \midrule
    (\textit{Model})                                                       & \multicolumn{4}{c}{\cellcolor[HTML]{E5E4E2} (b) \textbf{Ising 2D}}                                                                                                                                                                               \\
    $10.0$                                                                 & {$-0.1586$} (\textcolor{gray}{\textit{$-0.1586$}})                 & {$-0.1645$} (\textcolor{gray}{\textit{$-0.1610$}})      & {$-0.1589$} (\textcolor{gray}{\textit{$-0.1614$}})      & {$-0.1587$} (\textcolor{gray}{\textit{$-0.1587$}})      \\
    $20.0$                                                                 & {$0.2688$} (\textcolor{gray}{\textit{$0.2692$}})                   & {$0.2663$} (\textcolor{gray}{\textit{$0.2672$}})        & {$0.2680$} (\textcolor{gray}{\textit{$0.2677$}})        & {$0.2730$} (\textcolor{gray}{\textit{$0.2688$}})        \\
    $30.0$                                                                 & {$0.7771$} (\textcolor{gray}{\textit{$0.7777$}})                   & {$0.7786$} (\textcolor{gray}{\textit{$0.7800$}})        & {$0.7799$} (\textcolor{gray}{\textit{$0.7797$}})        & {$0.7812$} (\textcolor{gray}{\textit{$0.7812$}})        \\
    $40.0$                                                                 & {$0.9635$} (\textcolor{gray}{\textit{$0.9635$}})                   & {$0.9641$} (\textcolor{gray}{\textit{$0.9651$}})        & {$0.9647$} (\textcolor{gray}{\textit{$0.9633$}})        & {$0.9654$} (\textcolor{gray}{\textit{$0.9635$}})        \\
    $50.0$                                                                 & {$0.9984$} (\textcolor{gray}{\textit{$0.9984$}})                   & {$0.9979$} (\textcolor{gray}{\textit{$0.9982$}})        & {$0.9982$} (\textcolor{gray}{\textit{$0.9983$}})        & {$0.9985$} (\textcolor{gray}{\textit{$0.9985$}})        \\
    $60.0$                                                                 & {$0.9980$} (\textcolor{gray}{\textit{$0.9979$}})                   & {$0.9978$} (\textcolor{gray}{\textit{$0.9981$}})        & {$0.9981$} (\textcolor{gray}{\textit{$0.9965$}})        & {$0.9986$} (\textcolor{gray}{\textit{$0.9984$}})        \\
    \midrule
    (\textit{Model})                                                       & \multicolumn{4}{c}{\cellcolor[HTML]{E5E4E2} (c) \textbf{LMG}}                                                                                                                                                                                    \\
    $30.0$                                                                 & $\mathbf{0.4775}$ (\textcolor{gray}{\textit{$0.4774$}})            & $\mathbf{0.4762}$ (\textcolor{gray}{\textit{$0.4729$}}) & $\mathbf{0.4766}$ (\textcolor{gray}{\textit{$0.4770$}}) & $\mathbf{0.4776}$ (\textcolor{gray}{\textit{$0.4774$}}) \\
    $60.0$                                                                 & $\mathbf{0.5828}$ (\textcolor{gray}{\textit{$0.5828$}})            & $\mathbf{0.5792}$ (\textcolor{gray}{\textit{$0.5803$}}) & $\mathbf{0.5818}$ (\textcolor{gray}{\textit{$0.5815$}}) & $\mathbf{0.5828}$ (\textcolor{gray}{\textit{$0.5828$}}) \\
    $100.0$                                                                & $\mathbf{0.7471}$ (\textcolor{gray}{\textit{$0.7467$}})            & $\mathbf{0.7447}$ (\textcolor{gray}{\textit{$0.7468$}}) & $\mathbf{0.7459}$ (\textcolor{gray}{\textit{$0.7460$}}) & $\mathbf{0.7472}$ (\textcolor{gray}{\textit{$0.7471$}}) \\
    $200.0$                                                                & $\mathbf{0.8591}$ (\textcolor{gray}{\textit{$0.8592$}})            & $\mathbf{0.8561}$ (\textcolor{gray}{\textit{$0.8568$}}) & $\mathbf{0.8583}$ (\textcolor{gray}{\textit{$0.8587$}}) & $\mathbf{0.8591}$ (\textcolor{gray}{\textit{$0.8591$}}) \\
    $300.0$                                                                & $\mathbf{0.9093}$ (\textcolor{gray}{\textit{$0.9098$}})            & $\mathbf{0.9091}$ (\textcolor{gray}{\textit{$0.9077$}}) & $\mathbf{0.9095}$ (\textcolor{gray}{\textit{$0.9093$}}) & $\mathbf{0.9101}$ (\textcolor{gray}{\textit{$0.9104$}}) \\
    $500.0$                                                                & $\mathbf{0.9312}$ (\textcolor{gray}{\textit{$0.9368$}})            & $\mathbf{0.9349}$ (\textcolor{gray}{\textit{$0.9363$}}) & $\mathbf{0.9367}$ (\textcolor{gray}{\textit{$0.9367$}}) & $\mathbf{0.9372}$ (\textcolor{gray}{\textit{$0.9371$}}) \\
    $1000.0$                                                               & $\mathbf{0.9463}$ (\textcolor{gray}{\textit{$0.9484$}})            & $\mathbf{0.9522}$ (\textcolor{gray}{\textit{$0.9475$}}) & $\mathbf{0.9505}$ (\textcolor{gray}{\textit{$0.9510$}}) & $\mathbf{0.9518}$ (\textcolor{gray}{\textit{$0.9523$}}) \\
    $1500.0$                                                               & $\mathbf{0.9474}$ (\textcolor{gray}{\textit{$0.9436$}})            & $\mathbf{0.9385}$ (\textcolor{gray}{\textit{$0.9336$}}) & $\mathbf{0.9444}$ (\textcolor{gray}{\textit{$0.9524$}}) & $\mathbf{0.9499}$ (\textcolor{gray}{\textit{$0.9453$}}) \\
    $2000.0$                                                               & $\mathbf{0.9447}$ (\textcolor{gray}{\textit{$0.9448$}})            & $\mathbf{0.9646}$ (\textcolor{gray}{\textit{$0.9496$}}) & $\mathbf{0.9648}$ (\textcolor{gray}{\textit{$0.9521$}}) & $\mathbf{0.9636}$ (\textcolor{gray}{\textit{$0.9562$}}) \\
    \bottomrule
  \end{tabular}
  \caption{\textbf{Energy ratio obtained by \agentname}: \agentname using the Hamiltonian pool without identity (\textbf{bold}, see Appendix~\ref{sec:model}) and  with identity operator (\textcolor{gray}{gray} in the parenthesis, see Appendix~\ref{sec:identity_model}). Sector (a): 1D spin-$1/2$ Ising chain ($N\!=\!8$); Sector (b): 2D spin-$1/2$ Ising chain ($N\!=\!3\times3$); Sector (c): LMG model ($N\!=\!100$) at $h/J=0.9$.  We use $\gamma=0.1$ for Gaussian noise and $\delta=0.1$ for the gate rotation noise (see Appendix.~\ref{sec:noise}).}
  \label{tab:qmcts_fid}
\end{table}

\section{Additional numerical results on the energy landscape}

In Section~\ref{sec:landscape} we reported the discrete landscape of the generalized QAOA ansatz under the condition that the continuous variables are solved with high quality with the improved NPG solver. Here we include the landscape under another physical model,~i.e.~2D Ising model. We consider the case where $(\abs{\mathcal A}, q) \!=\! (5, 8)$, and the total number of gate sequences is thus $81920$. Similar to the plots displayed in Section~\ref{sec:landscape}, the landscape with a longer total duration ($JT=50$) features a dominant cluster at the rightmost part of the histogram. When the total duration is smaller, the number of clusters increases, and is shifted to the left.

\begin{figure}[!htbp]

  \includegraphics[width=1.1\textwidth]{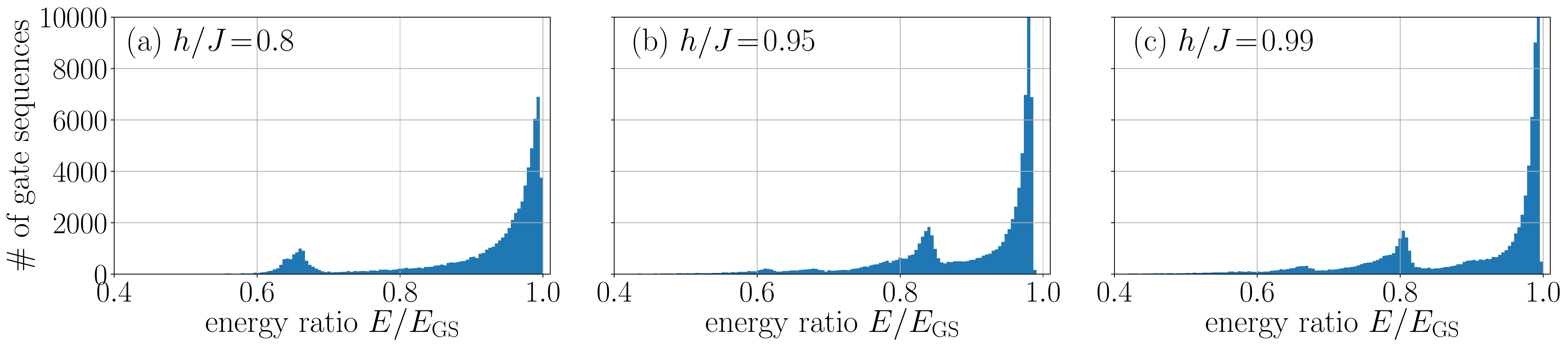}

  \caption{\textbf{Discrete landscape of LMG model with respect to different parameter $h/J$}: (a-c): Histograms of the energy ratio optimized by the improved natural gradient solver for $h/J= 0.8, 0.95$, and  $0.99$, respectively with gate duration $JT\!=\!1500$. $N_{\text{hist}}=81920$ samples are chosen from the discrete gate sequences of generalized QAOA with parameters $q=8$, $\abs{\mathcal A}=5$ and $N=100$. For the LMG model, the gap between the right-most peak and $1$ is larger when $h/J$ is between $0.8$ and $0.99$.}
  \label{fig:land_LMG_h}
\end{figure}

Figure~\ref{fig:land_LMG_h}  shows the influence of the parameter $h/J$ in the discrete landscape for the LMG model with gate duration $JT\!=\!1500$ and $N\!=\!100$. When $h/J\!=\!0.8$ and $h/J\!=\!0.99$, the rightmost peak in the energy ratio histogram gets close to $1$, which means that reaching the ground state would be a easy task in these two cases. The more difficult cases lies in between, for example when $h/J=0.95$. For the parameter $h/J=0.9$ we choose in the main text, there is a bigger gap (cf.~Fig.~\ref{fig:landscape_LMG}(c)) between the rightmost peak of the energy ratio and $1$, which means the problem we choose to solve is relatively challenging.

\begin{figure}[!htbp]

  \includegraphics[width=1.1\textwidth]{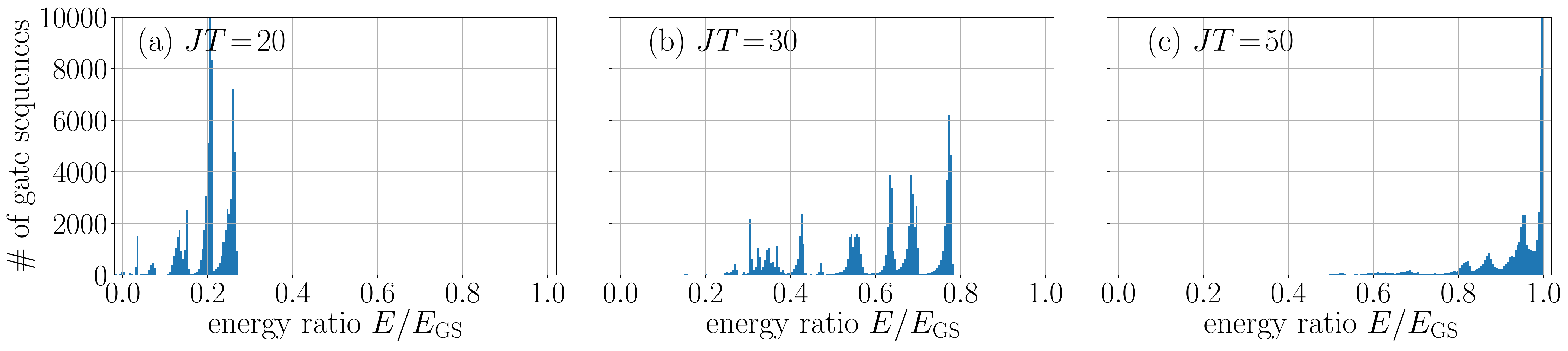}

  \caption{\textbf{Discrete landscape of 2D Ising model}: (a-c): Histograms of the energy ratio optimized by the improved natural gradient solver for $JT=20, 30, 50$, respectively. $N_{\text{hist}}=81920$ samples are chosen from the discrete gate sequences of generalized QAOA with parameters $q=8$ and $\abs{\mathcal A}=5$.}
  \label{fig:land_ising2d}
\end{figure}

\section{Physical models with the identity action}
\label{sec:identity_model}
The generalized QAOA ansatz provides us the freedom of adding different Hamiltonians to the Hamiltonian pool. One meaningful addition is the identity operator. Here, the identity operator corresponds to the identity gate that does not move the quantum state. If we take $\tilde H=\mathbf{0}$, then its corresponding unitary gate will be identity,~i.e.~$\exp (-i \tilde H \tilde \alpha )=\mathbf{I}$. This approach adds an extra amount of freedom to the optimization since the quantum control no longer needs to figure out how to \textit{exactly} distribute the gate duration budgets to different gates so as to reach the ground state. In other words, the original optimization problem (see Eqn.~\ref{eq:sim_problem}) becomes the relaxed form:
\begin{equation}
  \label{eq:sim_problem_relax}
  \min_{\{\alpha_j\}_{j=1}^q} \; \!\!\!\!\left\{ E(\{\alpha_j\}_{j=1}^q, \bt)\! :\! \sum_{j=1}^q \alpha_j \!\leq \! T;\ 0 \leq \alpha_j \leq T \right\}.
\end{equation}
With the identity action, the extended action space becomes $\mathcal{A}\!=\!\left\{
  \mathbf{0},
  \frac{H_1}{||H_1||},
  \frac{H_2}{||H_2||},
  \frac{A_1}{||A_1||},
  \frac{A_2}{||A_2||},
  \frac{A_3}{||A_3||}
  \right\},$ with the definitions shown in Appendix.~\ref{sec:model} for three different physics models.

\begin{figure}[!htbp]

  \includegraphics[width=1.0\textwidth]{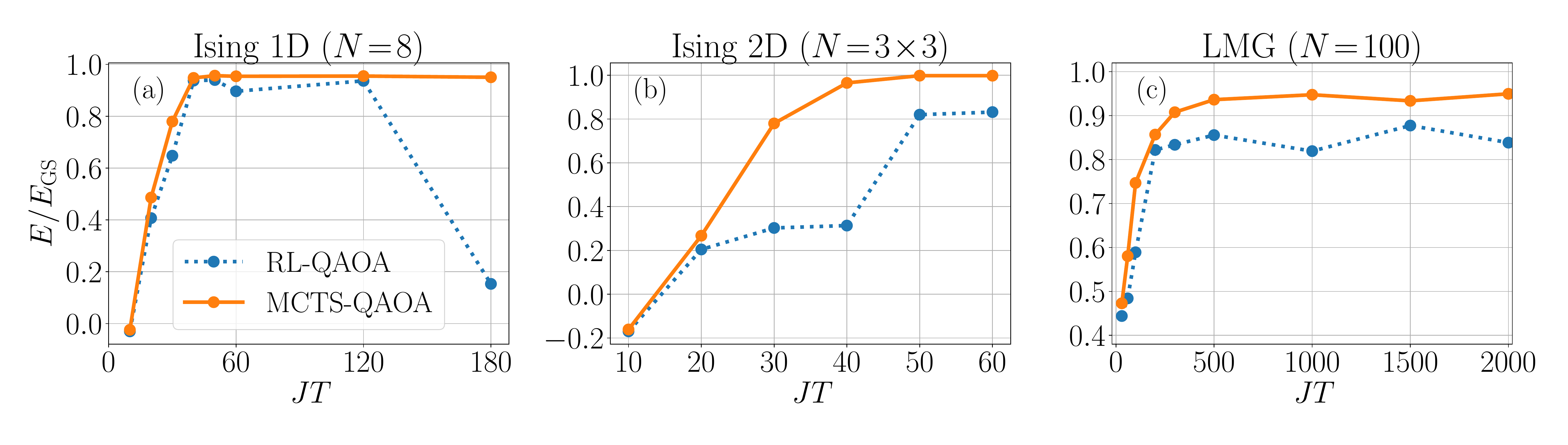}

  \includegraphics[width=1.0\textwidth]{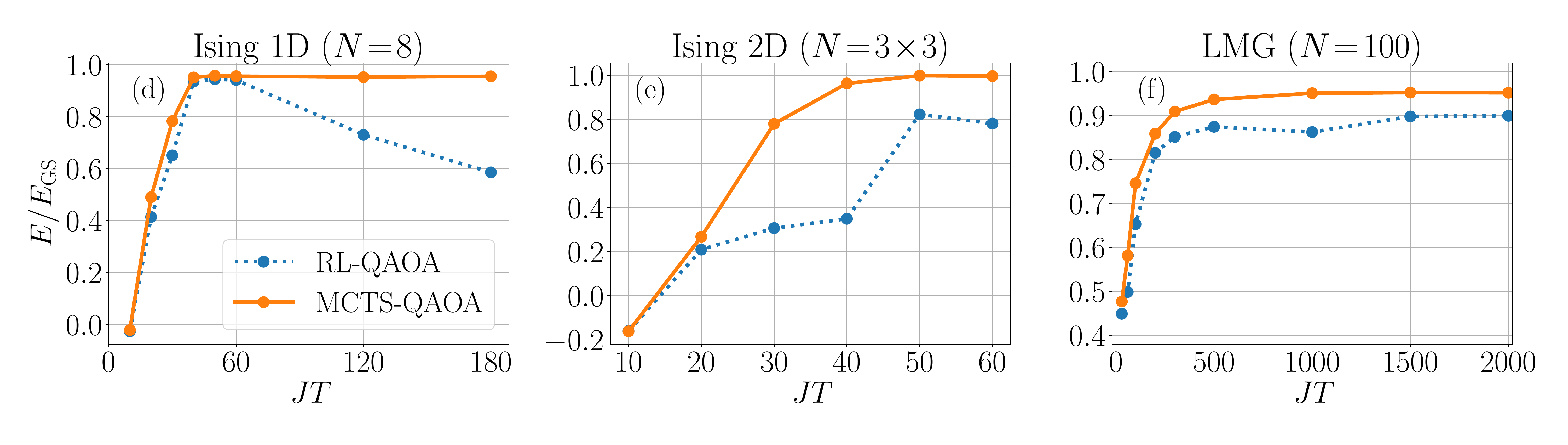}

  \includegraphics[width=1.0\textwidth]{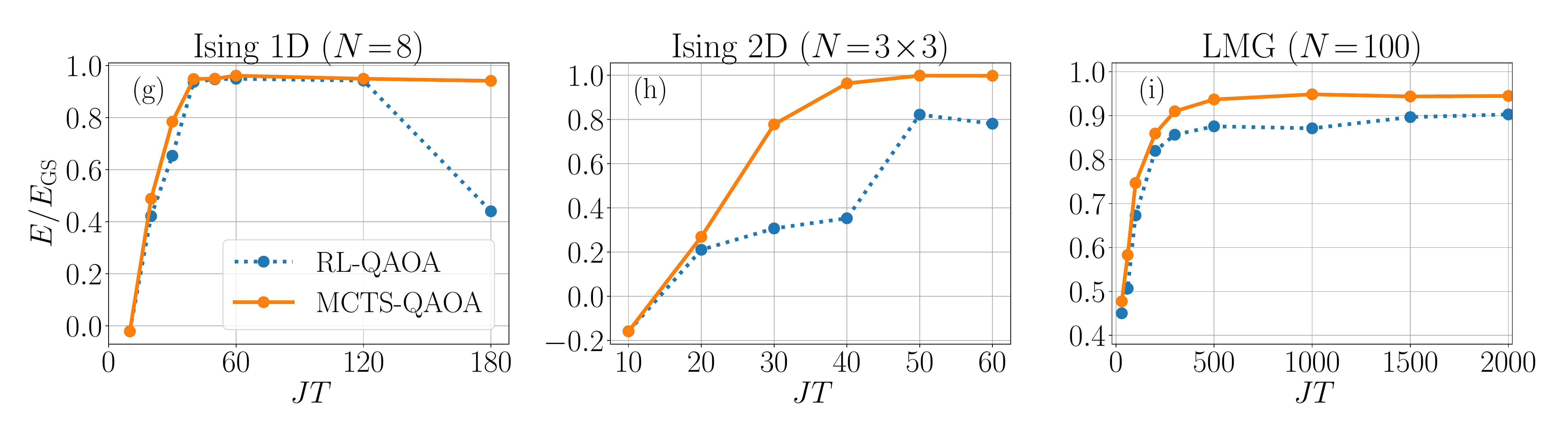}

  \includegraphics[width=1.0\textwidth]{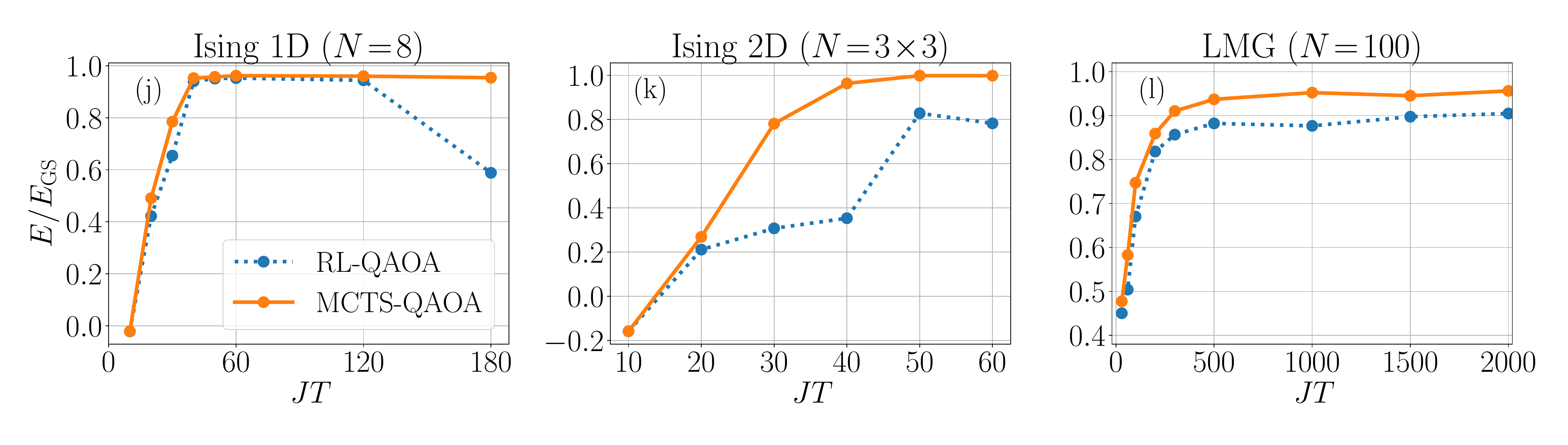}

  \vspace{-2em}
  \caption{Comparison between \agentname{} and RL-QAOA using the Hamiltonian pool with the identity operation. The physics setup is the same as that in Figure~\ref{fig:qmcts_quantum}. (a-c): quantum measurement noise; (d-f): Gaussian noise with $\gamma=0.1$; (g-i): gate rotation noise with $\delta=0.1$; (k-l): experiments without noise (cf. Appendix~\ref{sec:noise}). }
  \label{fig:qmcts_other_noise_identity}
\end{figure}

In this setting, a similar behavior is observed as in Figure~\ref{fig:qmcts_quantum} and Figure~\ref{fig:qmcts_other_noise}, which is shown in Figure~\ref{fig:qmcts_other_noise_identity}.
We conclude that \agentname{} outperforms RL-QAOA in all settings and \agentname still maintains a robust performance when that of RL-QAOA begins to deteriorate in the regime of large total gate durations. The raw data of the energy ratio obtained by \agentname{} is reported in Table~\ref{tab:qmcts_fid} (highlighted in gray), which also gives a direct comparison with the energy ratios obtained without the identity action.
It can be seen that the performance of \agentname{} in this setting is on par with the setting presented in the main text.

\end{document}